\documentclass[sigconf]{acmart}
\usepackage{algorithm}
\usepackage{algpseudocode}
\usepackage{bm}
\usepackage{enumitem}
\usepackage{multirow}
\usepackage{makecell}
\usepackage{pifont}
\usepackage{stfloats}

\AtBeginDocument{%
  }

\setcopyright{acmlicensed}

\copyrightyear{2025}
\acmYear{2025}
\acmDOI{XXXXXXX.XXXXXXX}

\acmConference[MICRO 2025]{The 58th IEEE/ACM International Symposium on Microarchitecture}{October 18--22, 2025}{Seoul, Korea}
\acmISBN{978-X-XXXX-XXXX-X/XX/XX}

\author{
    Minnan Pei\textsuperscript{1,2},
    Gang Li\textsuperscript{1\dag}, 
    Junwen Si\textsuperscript{1},
    Zeyu Zhu\textsuperscript{1,2},
    Zitao Mo\textsuperscript{1},
    Peisong Wang\textsuperscript{1,2},
    Zhuoran Song\textsuperscript{3},
    Xiaoyao Liang\textsuperscript{3},
    Jian Cheng\textsuperscript{1,2\dag} 
}

\affiliation{\institution{\textsuperscript{1}The Key Laboratory of Cognition and Decision Intelligence for Complex Systems, CASIA, Beijing, China}\city{}\country{}}
\affiliation{\institution{\textsuperscript{2}University of Chinese Academy of Sciences, Beijing, China}\city{}\country{}}
\affiliation{\institution{\textsuperscript{3}Shanghai Jiao Tong University, Shanghai, China}\city{}\country{}}

\renewcommand{\shortauthors}{Minnan Pei, et al.}
\begin{document}

\setcounter{page}{1} 

\title{GCC: A 3DGS Inference Architecture with Gaussian-Wise and Cross-Stage Conditional Processing}
% \subtitle{\normalsize{MICRO 2025 Submission
%     \textbf{\#538} -- Confidential Draft -- Do NOT Distribute!!}}
% \author{\normalsize{ISCA 2025 Submission
%    \textbf{\#NaN} -- Confidential Draft -- Do NOT Distribute!!}}



%%%%%% -- PAPER CONTENT STARTS-- %%%%%%%%

\begin{abstract}

3D Gaussian Splatting (3DGS) has emerged as a leading neural rendering technique for high-fidelity view synthesis, prompting the development of dedicated 3DGS accelerators for resource-constrained platforms. The conventional decoupled preprocessing-rendering dataflow in existing accelerators has two major limitations: 1) a significant portion of preprocessed Gaussians are not used in rendering, and 2) the same Gaussian gets repeatedly loaded across different tile renderings, resulting in substantial computational and data movement overhead. To address these issues, we propose GCC, a novel accelerator designed for fast and energy-efficient 3DGS inference. GCC introduces a novel dataflow featuring: 1) \textit{cross-stage conditional processing}, which interleaves preprocessing and rendering to dynamically skip unnecessary Gaussian preprocessing; and 2) \textit{Gaussian-wise rendering}, ensuring that all rendering operations for a given Gaussian are completed before moving to the next, thereby eliminating duplicated Gaussian loading. We also propose an alpha-based boundary identification method to derive compact and accurate Gaussian regions, thereby reducing rendering costs. We implement our GCC accelerator in 28nm technology. Extensive experiments demonstrate that GCC significantly outperforms the state-of-the-art 3DGS inference accelerator, GSCore, in both performance
and energy efficiency.

\end{abstract}
\begin{CCSXML}
<ccs2012>
   <concept>
       <concept_id>10010520.10010521</concept_id>
       <concept_desc>Computer systems organization~Architectures</concept_desc>
       <concept_significance>500</concept_significance>
       </concept>
   <concept>
       <concept_id>10010147.10010371.10010372.10010373</concept_id>
       <concept_desc>Computing methodologies~Rasterization</concept_desc>
       <concept_significance>300</concept_significance>
       </concept>
   <concept>
       <concept_id>10010583.10010600.10010615</concept_id>
       <concept_desc>Hardware~Logic circuits</concept_desc>
       <concept_significance>100</concept_significance>
       </concept>
   <concept>
       <concept_id>10010147.10010178.10010224</concept_id>
       <concept_desc>Computing methodologies~Computer vision</concept_desc>
       <concept_significance>300</concept_significance>
       </concept>
 </ccs2012>
\end{CCSXML}

\ccsdesc[500]{Computer systems organization~Architectures}
\ccsdesc[300]{Computing methodologies~Rasterization}
\ccsdesc[100]{Hardware~Logic circuits}
\ccsdesc[300]{Computing methodologies~Computer vision}

%
% Keywords. The author(s) should pick words that accurately describe
% the work being presented. Separate the keywords with commas.
\keywords{Accelerators, Gaussian Splatting, 3D Rendering}

\maketitle
\renewcommand{\thefootnote}{\dag}
\footnotetext{Co-corresponding authors.}
\renewcommand{\thefootnote}{\arabic{footnote}}

\section{Introduction}\label{sec:1}

In recent years, neural rendering techniques \cite{yu2021plenoctrees, fridovich2022plenoxels, mildenhall2021nerf,xu2022point, barron2022mip,kerbl20233d, wei2024emd, wei2025graphavatar} have advanced significantly in computer graphics and vision. Among these, 3D Gaussian Splatting (3DGS) \cite{kerbl20233d} has emerged as a leading method, combining the advantages of continuous volumetric representations with explicit point clouds to deliver high-quality, real-time novel-view synthesis on high-end GPUs. However, efficient deployment of 3DGS models on mobile devices like AR headsets \cite{li2022rt} remains challenging due to limited computational power and battery life. For instance, a comfortable immersive experience demands a rendering rate of at least 90 FPS—a target the original 3DGS algorithm, often operating below 10 FPS on mobile platforms, falls far short of \cite{kerbl20233d, huang2025seele}. This has spurred growing research in academia and industry on dedicated accelerators for fast and energy-efficient 3DGS inference \cite{lee2024gscore}.

Existing hardware accelerators \cite{lee2024gscore,wu2024gauspu,10946749} for 3DGS follow the standard processing pipeline of GPU implementation, featuring a \textit{two-stage ``preprocess-then-render'' process} and \textit{tile-wise rendering}. Specifically, for a given viewpoint, the millions of trained 3D Gaussians first undergo preprocessing: Gaussians outside the view frustum are culled, while the remaining ones are projected into 2D representations. These 2D Gaussians are then rendered tile-by-tile (e.g., 16$\times$16 pixels) to compute per-pixel RGB values. For each tile, all overlapping Gaussians are retrieved, sorted by depth, and processed via $\alpha$-blending to determine the final colors. 
While showing promise, we identify that this standard dataflow employed in existing 3DGS accelerators exhibits two major limitations, preventing them from achieving optimal performance and energy efficiency:

\begin{figure*}[t]
    \centering
    \includegraphics[width=\textwidth]{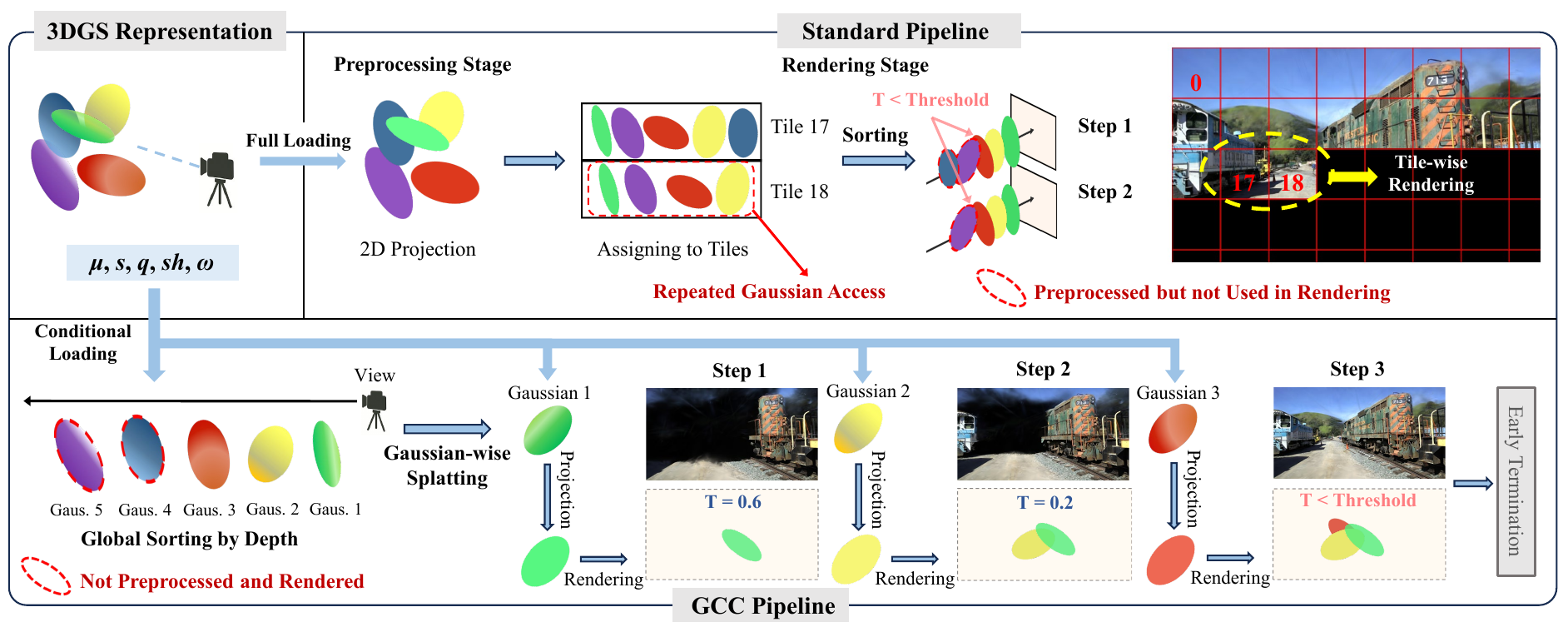}
    \vspace{-20pt}
    \caption{An illustration of the standard and our proposed dataflow for 3DGS inference. GCC exploits a cross-stage conditional processing scheme to dynamically eliminate redundant Gaussian preprocessing, and a Gaussian-wise approach to mitigate duplicated Gaussian accesses during rendering.}
    \vspace{-10pt}
    \label{fig:f2}
\end{figure*}

\textbf{First, the two-stage ``preprocess-then-render” approach introduces redundant Gaussian movements and computations.} During the rendering phase, the early termination strategy in $\alpha$-blending 
results in significantly fewer Gaussians are used in rendering than are preprocessed.
This implies that the 3D Gaussians corresponding to those non-rendered 2D ones can be entirely excluded during preprocessing, thereby reducing both data movements and computations. However, in the standard two-stage dataflow, all 3D Gaussians undergo preprocessing regardless of their actual utilization in subsequent rendering, introducing substantial redundancy. For example, our profiling reveals that approximately 40\% of runtime in the state-of-the-art GSCore~\cite{lee2024gscore} accelerator is dedicated to preprocessing 3D Gaussians, while over 60\% of the preprocessed Gaussians remain unused during subsequent rendering.

\textbf{Second, the tile-wise rendering incurs repeated data accesses for the same Gaussians.} At the beginning of rendering, all preprocessed 2D Gaussians establish indexing relationships with image tiles based on their spatial coordinates and bounding sizes. The rendering then processes tiles in scanline order: for each tile, it 1) retrieves all associated Gaussians, 2) sorts them by depth, and 3) computes $\alpha$-blending to determine RGB values for every pixel within the tile. For tiles of fixed size, a Gaussian overlapping multiple tiles is repeatedly loaded for each tile's rendering. This leads to total Gaussian loads far exceeding the number of unique Gaussians, causing significant memory inefficiency.

To address these issues, this paper proposes GCC, a specialized accelerator designed for fast and energy-efficient 3DGS inference. It introduces a novel dataflow that incorporates two key techniques: \textit{cross-stage conditional processing} and \textit{Gaussian-wise rendering}. 
On the one hand, unlike the standard pipeline, which strictly separates Gaussian preprocessing and rendering into two sequential stages, we interleave them conditionally at runtime, preprocessing and rendering only the Gaussians that contribute to the final RGB values. This avoids redundant Gaussian movements and computations present in the preprocessing stage of the conventional two-stage dataflow. On the other hand, to mitigate the issue of repeatedly loading the same Gaussian in different tiles during tile-wise rendering, we adopt a Gaussian-wise rendering approach. This involves rendering the complete Gaussians one by one in depth order, enabling each Gaussian’s feature data to be loaded only once throughout rendering. To further reduce rendering cost, we propose an alpha-based boundary identification method to derive compact and accurate Gaussian regions. In summary, the contributions of this paper are as follows:
\begin{itemize}
    \item We propose a novel 3DGS inference dataflow that integrates two key techniques: cross-stage conditional processing and Gaussian-wise rendering. This approach can dynamically eliminate both preprocessing and rendering operations for ineffective Gaussians, while reducing the overhead of repeated loading of the same Gaussians in conventional tile-wise rendering.
    \item We propose an alpha-based boundary identification method to dynamically determine the effective regions of each Gaussian during rendering. Combining expansion-based alpha computation with early termination, it can yield tighter Gaussian boundaries, reducing rendering costs while preserving accuracy.
    \item We propose GCC, a 3DGS inference accelerator tailored for the proposed dataflow. Evaluations show that GCC can achieve an average of 5.24$\times$ and 3.35$\times$ area-normalized speedup and energy efficiency, respectively, over state-of-the-art 3DGS inference accelerator GSCore, with a peak throughput of 667 FPS on the Lego dataset.
\end{itemize}

The rest of this paper is organized as follows. Section~\ref{sec:2} presents background and motivates the need for a more efficient dataflow and architecture. Section~\ref{sec:3} introduces the proposed GCC dataflow. Section~\ref{sec:4} details the hardware architecture of GCC. Section~\ref{sec:5} provides evaluation results. Section~\ref{sec:6} briefly introduces related work, and Section~\ref{sec:7} concludes this paper.

\section{Background and Motivation}\label{sec:2}

\subsection{3D Gaussian Splatting}

3D Gaussian Splatting~\cite{kerbl20233d}, as an emerging explicit scene representation approach, has drawn considerable attention due to its high-performance rendering capabilities~\cite{bao20253d, wu2024recent, fei20243d, jiang2024vr, dalal2024gaussian}. 3DGS represents scenes as a set of anisotropic Gaussian points continuously distributed in space, directly rasterizing these Gaussian points (i.e., splatting) to synthesize novel views. Compared to traditional implicit NeRF approaches~\cite{mildenhall2021nerf} (e.g., using deep MLPs), 3DGS explicitly represents scene elements and avoids per-pixel neural network inference, achieving significant improvements in rendering speed up to real-time frame rates desktop-class GPU~\cite{kerbl20233d} ($\geq$ 30 FPS at 1080p). It simultaneously maintains the differentiability advantages of NeRF and the sparsity benefits of explicit point clouds.

The rendering pipeline of 3DGS employs the classical EWA splatting method~\cite{zwicker2002ewa} but incorporates several critical optimization strategies for efficient GPU acceleration, including 1) data compression and memory-friendly representations, 2) preprocessing and rendering optimizations, and 3) early termination.
First, to reduce storage overhead and memory bandwidth demands, 3DGS compresses Gaussian parameters into a compact representation, each consisting of position coordinates, spherical harmonics (SH) for color, and covariance matrices. Specifically, position coordinates are represented as 3D spatial vectors, colors as 48 spherical harmonic coefficients, and covariance matrices through scale factors and rotation quaternion factors, accompanied by a transparency parameter~\cite{sloan2023precomputed}. Each Gaussian point has its unique set of parameters, with camera parameters varying per viewpoint, including camera poses, view matrices, and projection matrices.
Second, 3DGS optimizes the rendering pipeline for GPU characteristics, transforming the 3D world coordinate system first into a normalized device coordinate (NDC) space and then into the camera perspective. Major computational workloads involve numerous small matrix multiplications for projection ~\eqref{eq:1} and summation of spherical harmonic functions via Legendre polynomials ~\eqref{eq:2}.
\begin{equation}
\begin{aligned}
\Sigma &= R S S^{T} R^{T} 
,\quad
\Sigma' = J W \Sigma W^{T} J^{T}
\end{aligned}
\label{eq:1}
\tag{1}
\end{equation}
\begin{equation}
C_{RGB} = \sum_{l=0}^{L} \sum_{m=0}^{B_l - 1} c_{lm} \cdot f_{l,m}(x, y, z) \cdot sh_{b_l + m}
\label{eq:2}
\tag{2}
\end{equation}
Next, the preprocessing pipeline assigns an ellipse and an axis-aligned bounding box (AABB) to each Gaussian projected onto 2D, generating Gaussian-tile key-value pairs through tile coverage traversal. These Gaussians undergo sorting based on view depth using radix sorting, ensuring the blending will be correct order. Rasterization within each tile processes all pixels concurrently to compute each Gaussian’s current-layer transparency:
\begin{equation}
\alpha = \min\left(0.99, \, \omega \cdot e^{-\frac{1}{2} \cdot \bm{d}^T \Sigma'^{-1} \bm{d}} \right)
\tag{3}\label{eq:3}
\end{equation}
Final process updates transmittance and blending colors in depth-order~\cite{levoy1988display} shown as Equation~\eqref{eq:4} . To mitigate numerical instability during training, 3DGS excludes pixels in bounding box with alpha values less than 1/255 in forward and backward passes, and training terminates once cumulative transparency reaches 0.0001~\cite{kerbl20233d}. These termination criteria introduce predictable patterns in intermediate rendering variables, offering opportunities for computational efficiency improvements.
\begin{equation}
T_i = \prod_{j=1}^{i-1} (1 - \alpha_j), \quad 
C = \sum_{i=1}^{N} T_i \alpha_i c_i
\label{eq:4}
\tag{4}
\end{equation}

\begin{figure}[t]
    \centering
    \includegraphics[width=0.235\textwidth]{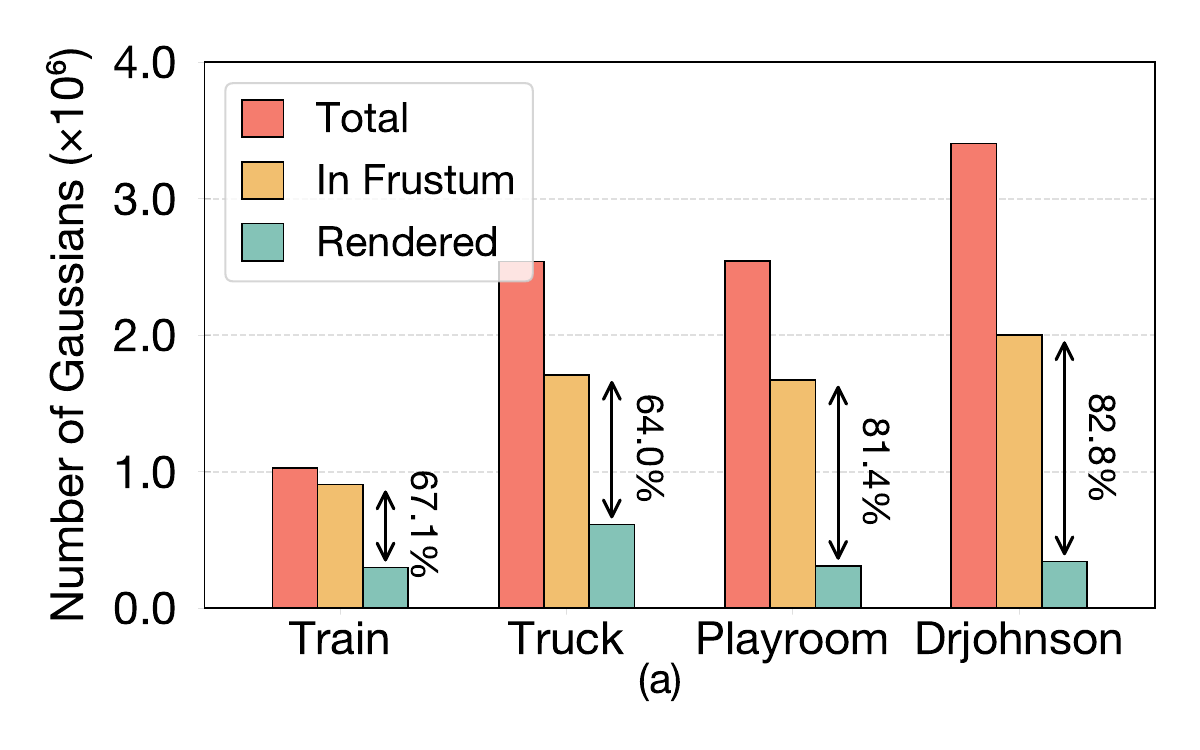}
    \hfill
    \includegraphics[width=0.235\textwidth]{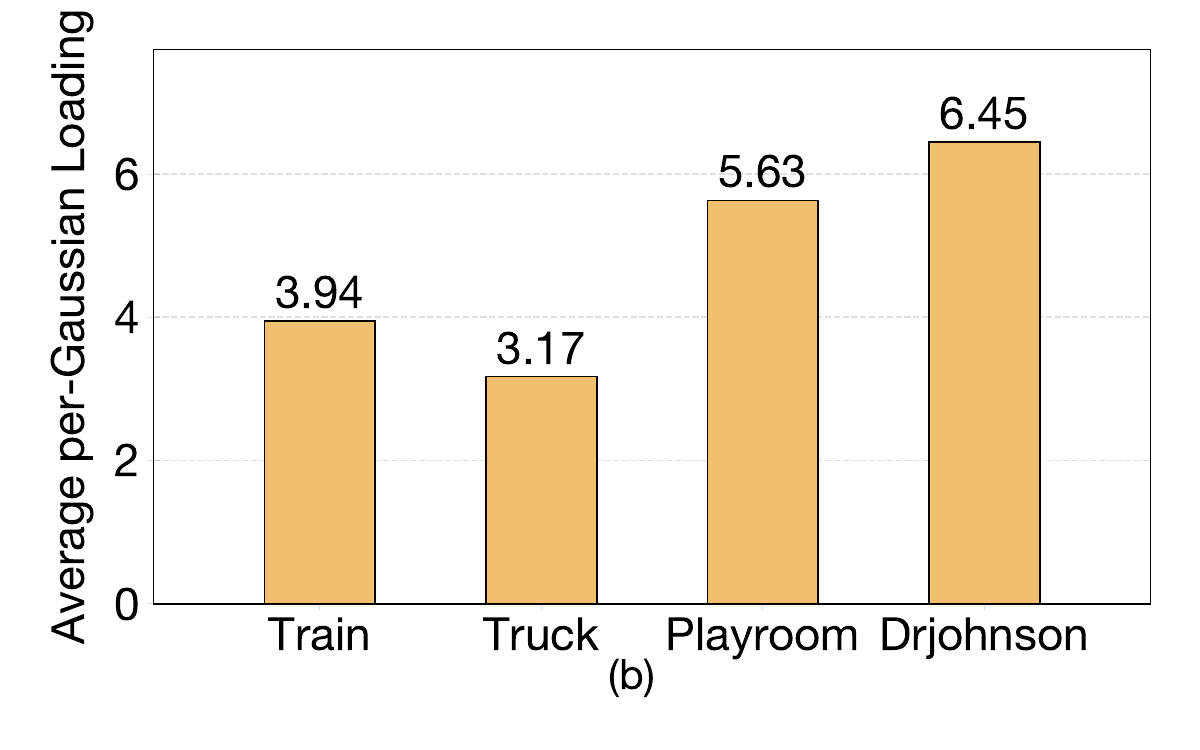}
    \vspace{-20pt}
    \caption{(a) The number of Gaussians in different processing phases; (b) The average number of per-Gaussian loadings during rendering in GSCore \cite{lee2024gscore}.}
    \vspace{-10pt}
    \label{fig:f1}
\end{figure}

\subsection{Challenges in 3DGS Accelerator Design}

Although 3DGS achieves faster inference speeds than NeRF \cite{mildenhall2021nerf} on high-performance GPUs, its deployment on resource-constrained mobile devices remains challenging~\cite{bao20253d}. To enhance speed and reduce energy consumption, several dedicated accelerators for 3DGS have recently been proposed, including: the first 3DGS inference architecture, GSCore \cite{lee2024gscore}; the 3DGS-based SLAM architecture, GauSPU \cite{wu2024gauspu}; and GSArch \cite{10946749}, an accelerator optimized for 3DGS training.

While promising, we observe that existing 3DGS accelerators adopt the same dataflow as GPU implementations (termed the ``standard dataflow” for clarity), characterized by two key features: 1) a two-stage ``preprocess-then-render” process, and 2) tile-wise rendering, as shown in Figure~\ref{fig:f2}. Specifically, in the preprocessing stage, the millions of trained 3D Gaussians are first culled to remove those outside the view frustum for a given viewpoint. They are then transformed into 2D representations through a series of matrix operations. In the rendering stage, each 2D Gaussian determines its boundaries using a specific method and establishes an indexed relationship with multiple fixed-size tiles on the image based on their overlap regions. The image is then rendered in scanline order, starting from the first tile. The rendering process for each tile includes operations such as Gaussian sorting and $\alpha$-blending. Through in-depth analysis, we identify two major limitations in this standard dataflow that hinder optimal hardware efficiency:

\emph{\textbf{Challenge 1: Decoupling preprocessing and rendering incurs significant redundancy in preprocessing.}} 
The standard 3DGS dataflow begins with a preprocessing stage, where each 3D Gaussian is independently projected into 2D representation through position transformation and color computation based on third-order spherical harmonics (SH). These 2D Gaussians then serve as input for the subsequent rendering stage. Nevertheless, owing to the inherent early termination mechanism in $\alpha$-blending during rendering, a significant portion (over 60\%, as shown in Figure~\ref{fig:f1} (a)) of the preprocessed 2D Gaussians are ultimately unused. Particularly, each 3D Gaussian is represented by 59 floating-point parameters, among which a staggering 81.4\% (48 out of 59) of the SH coefficients remain unused before $\alpha$-blending begins. Existing 3DGS accelerators, however, uniformly load and process all 3D Gaussians during preprocessing without considering their subsequent utilization in rendering. This leads to substantial inefficiencies in both off-chip data transfer and on-chip computation. For instance, GSCore \cite{lee2024gscore} spends up to 40\% of the total time on preprocessing, yet over half of the 2D Gaussians are discarded in the rendering stage. 

\emph{\textbf{Challenge 2: Tile-wise rendering causes duplicated loadings for Gaussians overlapping multiple tiles.}} In the standard dataflow, image rendering is performed at the tile granularity. The 2D Gaussians are mapped to several fixed-size tiles in the image based on their size and positional information, with larger Gaussians overlapping more tiles. For each tile’s rendering, all overlapping Gaussians are first sorted by depth. Their colors are then computed and blended from near to far using $\alpha$-blending to produce the tile’s final color. Evidently, when a Gaussian overlaps multiple tiles, its parameters should be loaded repeatedly during rendering due to limited on-chip storage. As shown in Figure~\ref{fig:f1}(b), we observe that in various tasks, the same Gaussian is loaded 3.17 to 6.45 times on average during rendering in GSCore \cite{lee2024gscore}. Therefore, although the number of Gaussians used in rendering is significantly lower than in preprocessing, tile-wise rendering still incurs substantial data transfer overhead from repeated Gaussian loading, leading to energy inefficiency.

\subsection{Our Approach} 

To address the above limitations of the standard dataflow and enable fast and energy-efficient 3DGS inference, we propose GCC, a novel inference accelerator with two key dataflow innovations:

\emph{\textbf{Cross-stage conditional processing.}}
Unlike the conventional decoupled pipeline, we propose a novel dataflow where Gaussian preprocessing and rendering are executed alternately. Once the rendering termination condition is met, all subsequent preprocessing and rendering operations for remaining Gaussians are skipped, thereby effectively eliminating redundant computations and data movements caused by unnecessary Gaussian preprocessing.

\emph{\textbf{Gaussian-wise rendering.}}
To avoid repeated off-chip loading of the same Gaussian in tile-wise rendering, we present a Gaussian-wise approach. Starting from the nearest Gaussian, we complete the entire rendering of one Gaussian before proceeding to the next. In this scheme, each Gaussian's parameters are loaded only once, significantly reducing data movement overhead.

As shown in Figure~\ref{fig:f2}, our proposed dataflow fundamentally unifies preprocessing and rendering into a Gaussian-wise processing pipeline, effectively mitigating the computational and data movement redundancies caused by the mismatch between Gaussian-wise preprocessing and tile-wise rendering in standard dataflow.

\begin{figure*}[t]
    \centering
    \includegraphics[width=\textwidth]{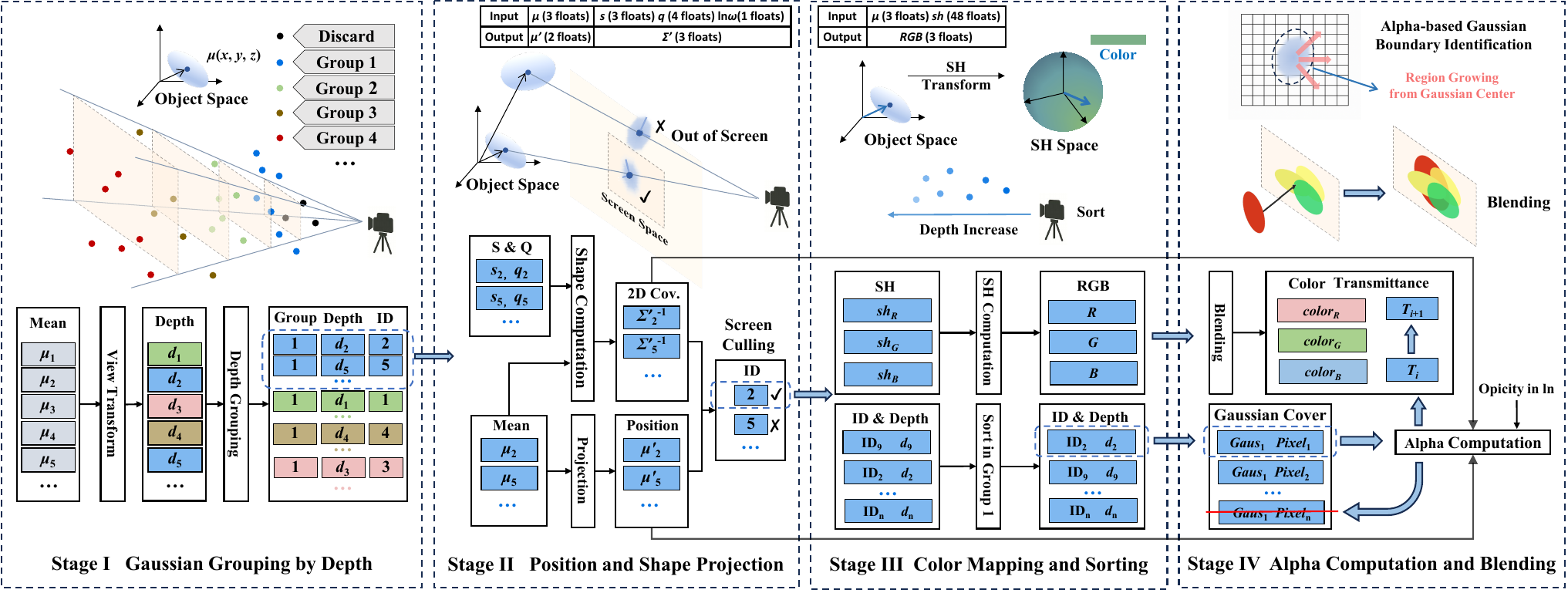} 
    \vspace{-20pt}
    \caption{Overview of GCC dataflow.}
    \vspace{-10pt}
    \label{fig:f4}
\end{figure*}

\section{GCC Dataflow}\label{sec:3}

Our GCC accelerator adopts a fundamentally different dataflow based on two principles: Gaussian-wise rendering and cross-stage conditional processing.
Unlike conventional 3DGS accelerators that statically separate preprocessing and rendering, our design treats each Gaussian as an independent compute primitive and dynamically triggers its processing only when necessary.

Figure~\ref{fig:f4} presents an overview of our dataflow. The pipeline consists of four stages: 1) Gaussian grouping by depth, 2) position and shape projection, 3) color mapping and sorting, and 4) alpha computation and blending. Each stage progressively filters out irrelevant Gaussians based on visibility, depth, or contribution, ensuring that only effective Gaussians participate in downstream computation. Notably, the initial grouping stage establishes the foundation for Gaussian-wise scheduling by organizing Gaussians in depth order, enabling early termination across groups. 

\subsection*{Stage I: Gaussian Grouping by Depth}

The GCC dataflow begins by computing the view-space depth ($\boldsymbol{d}$) of each Gaussian, based on its 3D mean position $\boldsymbol{\mu}$. This is achieved by applying a view matrix transformation to obtain $\boldsymbol{\mu}' = (x', y', z')$, where $z'$ is used as the depth value.
Gaussians are then assigned into depth bins according to their $\boldsymbol{d}$ values, forming multiple processing groups sorted from near to far. Those with $\boldsymbol{d}$ below a visibility threshold (e.g., 0.2) are immediately culled, as they fall outside the effective frustum. This early-stage Gaussian-wise culling reduces unnecessary projection and blending computations for invisible objects.
These depth groups not only dictate the order of execution but also enable coarse-grained conditional skipping: once the accumulated transmittance meets the early termination criteria, deeper groups are bypassed entirely. 

With such organization, Gaussians can be processed in a globally sorted order, eliminating the need for per-tile sorting and assignment in tile-wise pipelines, where each Gaussian must be redundantly sorted and associated with multiple tiles.

\subsection*{Stage II: Position and Shape Projection}

In this stage, each Gaussian in a depth group undergoes two concurrent transformations: position projection and shape projection.
First, the 3D mean $\boldsymbol{\mu}$ is transformed into 2D pixel coordinates $\boldsymbol{\mu'}$ via the camera projection matrix and normalized to screen space. This defines the center of the Gaussian's 2D influence region.
The scale vector $\boldsymbol{s}$ and rotation quaternion $\boldsymbol{q}$ are used to reconstruct the 3D covariance matrix $\boldsymbol{\Sigma}$, which is then projected into 2D as $\boldsymbol{\Sigma'}$ via a Jacobian matrix $\boldsymbol{J}$. The resulting matrix $\boldsymbol{\Sigma'}$ describes the elliptical footprint of the Gaussian under the current view.
The spatial extent of this footprint is typically approximated using a $3\sigma$ envelope, defined by the quadratic form:
\begin{equation}
\begin{aligned}
(\boldsymbol{p} - \boldsymbol{\mu'})^\top \boldsymbol{\Sigma'}^{-1} (\boldsymbol{p} - \boldsymbol{\mu'}) = 3^2
\end{aligned}
\label{eq:5}
\tag{5}
\end{equation}
where $\boldsymbol{p}$ is a 2D pixel position on the image plane. Equation~\ref{eq:5} represents the elliptical boundary within which the Gaussian has non-negligible influence. The eigenvalues $\lambda_1$ and $\lambda_2$ of $\boldsymbol{\Sigma'}$ correspond to the squared lengths of the major and minor axes of the ellipse, from which we derive the conservative bounding-box radius:
\begin{equation}
\begin{aligned}
r = \lceil 3 \cdot \sqrt{\max(\lambda_1, \lambda_2)} \rceil
\end{aligned}
\label{eq:6}
\tag{6}
\end{equation}
If the resulting AABB lies entirely outside the screen bounds, the Gaussian is culled.

However, the $3\sigma$ rule provides a fixed approximation that ignores the actual opacity $\omega$ of each Gaussian. To tighten the culling criterion, we adopt an \textit{opacity-aware envelope} based on the minimum alpha threshold $\alpha \geq 1/255$. Recall that alpha is computed as Equation~\ref{eq:3}. To ensure that $\alpha \geq 1/255$, the elliptical support must satisfy:
\begin{equation}
\begin{aligned}
(\boldsymbol{p} - \boldsymbol{\mu'})^\top \boldsymbol{\Sigma'}^{-1} (\boldsymbol{p} - \boldsymbol{\mu'}) \leq 2 \cdot \ln(255 \cdot \omega)
\end{aligned}
\label{eq:7}
\tag{7}
\end{equation}
This leads to a tighter radius estimate that varies with the opacity:
\begin{equation}
\begin{aligned}
r = \left\lceil \sqrt{2 \cdot \ln(255 \cdot \omega) \cdot \max(\lambda_1, \lambda_2)} \right\rceil
\end{aligned}
\label{eq:8}
\tag{8}
\end{equation}
We refer to this dynamic thresholding scheme as the \textit{$\omega$-$\sigma$ law}, which adapts the Gaussian footprint to its actual visual contribution. Compared to the static $3\sigma$ rule, this opacity-aware culling removes more redundant Gaussians while preserving visual fidelity.
At the end of this stage, only visible Gaussians remain. Their projected positions $\boldsymbol{\mu'}$ and inverse covariance matrices $\boldsymbol{\Sigma'}^{-1}$ are passed to the next stage for color computation.

\begin{figure}[t]
    \centering
    \includegraphics[width=0.235\textwidth]{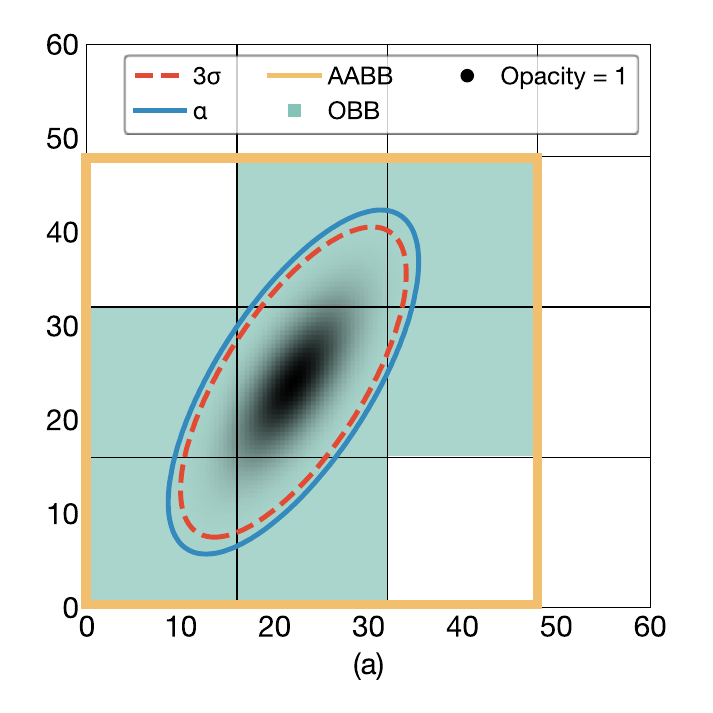}
    \hfill
    \includegraphics[width=0.235\textwidth]{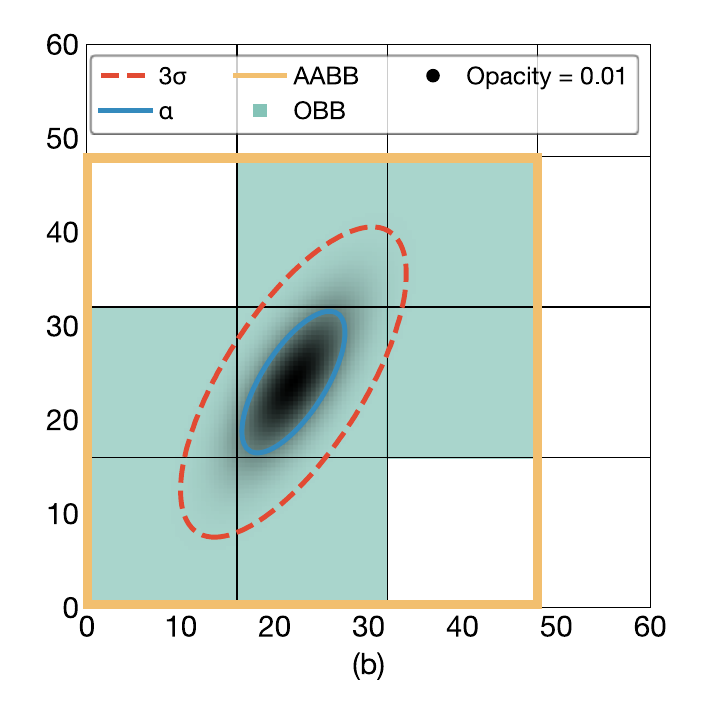}
    \vspace{-20pt}
    \caption{Different Gaussian regions under AABB, OBB methods, and the effictive regions (blue) affected by Opacity. (a) Opacity is 1; (b) Opacity is 0.01.}
    \vspace{-10pt}
    \label{fig:f3}
\end{figure}

\subsection*{Stage III: Color Mapping and Sorting}

In this stage, each Gaussian that survives visibility culling undergoes spherical harmonics (SH)–based color evaluation and intra-group depth sorting.
For color computation, we use third-order real spherical harmonics, where each RGB channel is represented by 16 coefficients, yielding 48 parameters per Gaussian. The normalized view direction $\boldsymbol{v} = (x, y, z)$ is used to evaluate the SH basis Equation \ref{eq:2}.
Meanwhile, we sort all Gaussians within each depth group in front-to-back order. Since blending in Stage IV is order-dependent, this step ensures that transparency is resolved correctly. The sorting is performed using the precomputed depth $\boldsymbol{d}$ values from Stage I.
By performing both color computation and sorting in a Gaussian-centric and group-local manner, this stage avoids global tile-based sorting, maintains processing regularity, and prepares each Gaussian for efficient $\alpha$-blending in the final stage.

\subsection*{Stage IV: Alpha Computation and Blending}

In the final stage, we perform alpha computation and blending to produce the final pixel values. This stage processes each Gaussian in front-to-back order, based on the sorted per-group list from Stage~III.

Each Gaussian casts its influence over a subset of pixels, determined by its projected position $\boldsymbol{\mu'}$ and 2D covariance matrix $\boldsymbol{\Sigma'}$ from Stage~II. For each pixel within this elliptical region, the alpha contribution is computed using the opacity $\omega$ and the offset from the projected center, following:
\begin{equation}\label{eq:9} \alpha = \min\left(0.99, e^{\ln\omega-\frac{1}{2} \cdot (\bm{p}'-\bm\mu')^T\Sigma'^{-1} (\bm{p}'-\bm\mu')} \right) \geq \frac{1}{255} \tag{9} \end{equation}
To avoid evaluating all pixels in the image for every Gaussian, we introduce Alpha-based Gaussian Boundary Identification, a runtime strategy that dynamically restricts each Gaussian’s evaluation to a minimal set of pixels based on its alpha footprint and a fixed perceptual threshold. This alpha-based identifier significantly reduces redundant weight calculations, especially for low-opacity or spatially compact Gaussians.

During compositing, we track the per-pixel accumulated transmittance $T$. Once the transmittance surpasses a predefined threshold, subsequent Gaussians are skipped to avoid redundant blending. Working in tandem, alpha-based identifier and early termination complete the final stage of our Gaussian-wise pipeline by eliminating both unnecessary region evaluation and downstream composition.

\begin{table}[t]
    \centering
    \caption{Average number of rendered pixels per frame.}
    \vspace{-5pt}
    \label{tab:t3}
    \resizebox{\linewidth}{!}{%
    \begin{tabular}{c | c c c c}
        \toprule
        \parbox{2.2cm}{\centering\textbf{Bounding-Box \\ Methods}} & 
        \parbox{1.5cm}{\centering\textbf{Train \\(M Pixels)}} & 
        \parbox{1.5cm}{\centering\textbf{Truck \\(M Pixels)}} & 
        \parbox{1.5cm}{\centering\textbf{Playroom \\(M Pixels)}} & 
        \parbox{1.5cm}{\centering\textbf{Drjohnson \\(M Pixels)}} \\
        \midrule
        AABB & 1164 & 1161 & 1177 & 1697 \\
        OBB & 378 & 416 & 333 & 460 \\
        Rendered & 31 & 32 & 60 & 73 \\
        \bottomrule
    \end{tabular}
    }
    \vspace{-15pt}
\end{table}

\subsection*{Alpha-based Gaussian Boundary Identification}

By transitioning tile-wise rendering to a Gaussian-wise paradigm that decouples computation from fixed tile boundaries, we must correspondingly establish an efficient method to delineate each Gaussian’s influence region in image space.
Although recent works such as GSCore \cite{lee2024gscore} adoptes Oriented Bounding Boxes (OBBs) and subtile partitioning to reduce redundancy over Axis-Aligned Bounding Boxes (AABBs), these strategies remain fundamentally tile-centric based on 3$\sigma$ criterion, failing to capture the true elliptical footprint of a Gaussian, which is governed by its alpha profile as defined in Equation~\ref{eq:9}. As shown in Figure~\ref{fig:f3} and Table~\ref{tab:t3}, many pixels outside a Gaussian’s actual influence are still processed in conventinal methods, with rendered pixel counts exceeding the effective region by 5–10$\times$ in typical scenes.

In contrast, our Gaussian-wise rendering paradigm enables per-Gaussian evaluation, naturally aligning with the elliptical influence defined by the alpha threshold. Leveraging this advantage, we introduce a runtime Alpha-based Boundary Identification technique (detailed in Algorithm~\ref{Algorithm1}) that dynamically determines the minimal set of contributing pixels for each Gaussian, eliminating redundant evaluations without compromising accuracy.
Unlike static bounding-box methods that often introduce substantial redundant computations, our approach dynamically identifies the Gaussian’s exact elliptical boundary in image space, guided by an alpha-based criterion established previously in Stage~II and Stage~IV.

The algorithm employs a breadth-first pixel traversal starting from the Gaussian's projected center $\boldsymbol{\mu'}$. If the projected center lies outside the image boundary, traversal initiates from the nearest valid pixel within image bounds. Pixels are explored progressively outward, and for each pixel visited, we evaluate the elliptical alpha condition $E(p)$ to determine whether it falls within the Gaussian’s influence region.
Crucially, due to the convexity of the Gaussian’s elliptical footprint, if a pixel fails the alpha condition, the algorithm can safely terminate further traversal in that direction by marking the corresponding pixel block as visited. This effectively skips unnecessary alpha evaluations across entire pixel blocks, substantially reducing redundant computations, especially around elliptical boundaries of low-opacity Gaussians.

\begin{algorithm}[h]
\caption{Alpha-based Gaussian Boundary Identification}
\label{Algorithm1}
\begin{algorithmic}[1]
\State \textbf{Input:} Pixel grid $P$, projected Gaussian center $\mu'$, effective condition $E(\cdot)$, visited map $S$
\State \textbf{Output:} Pixel set $P_{\text{influence}}$
\State $p_c \gets \text{FindNearestInBounds}(\mu', P)$
\State $Q \gets \text{new Queue()}$; $Q.push(p_c)$
\State $P_{\text{influence}} \gets \{p_c\}$
\State $S[p_c] \gets \text{true}$ \Comment{Initialize visited map}
\While{\textbf{not} $Q.empty()$}
    \State $p \gets Q.pop()$
    \For{each neighbor $q$ of $p$}
        \If{$q$ is in bounds \textbf{and} $S[q]$ is false}
            \State $S[q] \gets \text{true}$
            \If{$E(q)$}
                \State $P_{\text{influence}} \gets P_{\text{influence}} \cup \{q\}$
                \State $Q.push(q)$
            \EndIf
        \EndIf
    \EndFor
\EndWhile
\State \Return $P_{\text{influence}}$
\end{algorithmic}
\end{algorithm}

\section{GCC Architecture}\label{sec:4}

\subsection{Overview}

\begin{figure}
  \centering
  % \captionsetup{font=small}
\centering
    \includegraphics[width=\linewidth]{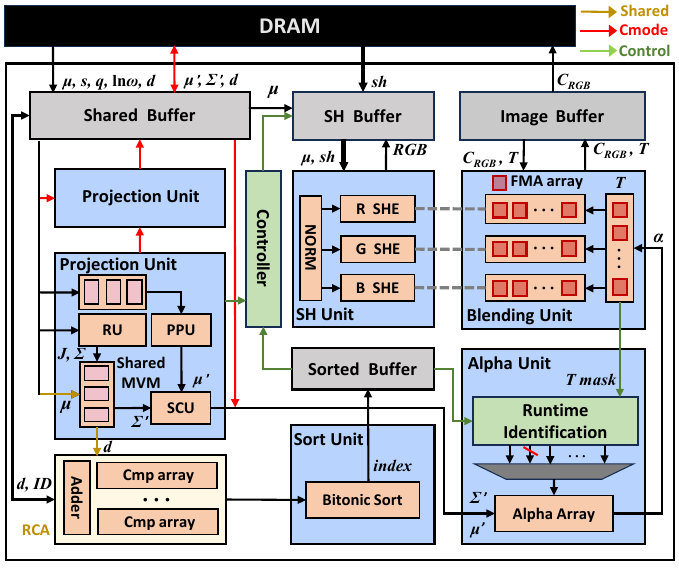}
    \vspace{-20pt}
    \caption{GCC Architecture.}
    \vspace{-10pt}
    \label{fig:f5}
\end{figure}

The overall GCC architecture, as shown in Figure ~\ref{fig:f5}, consists of a series of pipelined hardware modules: Reconfigurable Comparator Array (RCA), Projection Unit, Spherical Harmonics (SH) Unit, Sort Unit, Alpha Unit, and Blending Unit. Each module performs its core computations using floating-point fused multiply-add (FMA)~\cite{mach2020fpnew} operations to improve arithmetic efficiency.
The on-chip memory consists of several dedicated buffers. The Shared Buffer and SH Buffer handle data transfers of Gaussian parameters between DRAM and the compute pipeline. The Image Buffer stores intermediate results for color and transmittance accumulation. A local Sort Buffer temporarily holds sorted indices to avoid off-chip accesses.
Meanwhile, the entire rendering process is coordinated by an on-chip controller, which schedules execution stages based on depth grouping and sorting outcomes.

During processing, Gaussian model parameters and camera view data are streamed from off-chip memory. At the beginning of each frame, the Shared Matrix-Vector Multiplier (MVM) in the Projection Unit computes the depth of Gaussians by processing their 3D means in batches. These depth values are then used to group Gaussians into depth-aware bins.
For each Gaussian group, the 3D mean vectors $\mu$ are projected onto the 2D screen space using the Position Projection Unit (PPU). Meanwhile, shape parameters $\mu, s$ and $q$ are reconstructed by the Reconstruction Unit (RU) and converted into 2D inverse covariance matrices through the Shared MVM. A dedicated Screen Culling Unit (SCU) then removes Gaussians falling outside the current view frustum.
For each group of in-frustum Gaussians, the Sort Unit determines the rendering order using a 16-element bitonic sorting network, following the design in GSCore~\cite{lee2024gscore}. Once sorted, the spherical harmonics 
coefficients of each Gaussian are streamed into the Spherical Harmonics Element (SHE), where RGB colors are computed separately for each channel.
In the Alpha Unit, Gaussians are processed in pixel blocks using a alpha-based identifier algorithm to reduce unnecessary alpha computations. This block-wise execution aligns with the architecture’s processing array and ensures efficient memory access patterns. The valid transmittance weights and RGB contributions are then accumulated in the Blending Unit. Final results are written to the Image Buffer for subsequent output.

\begin{figure}[t]
    \centering
    \includegraphics[width=0.235\textwidth]{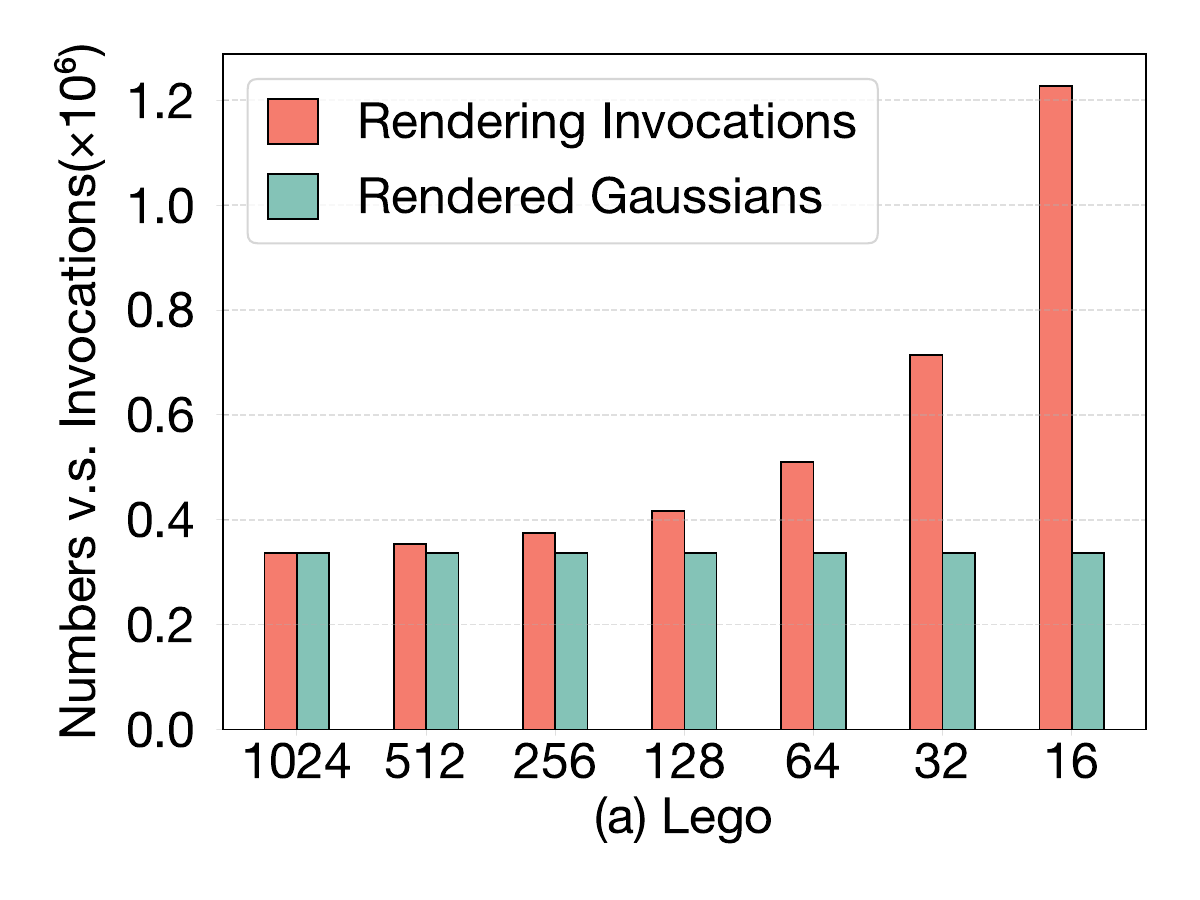}
    \hfill
    \includegraphics[width=0.235\textwidth]{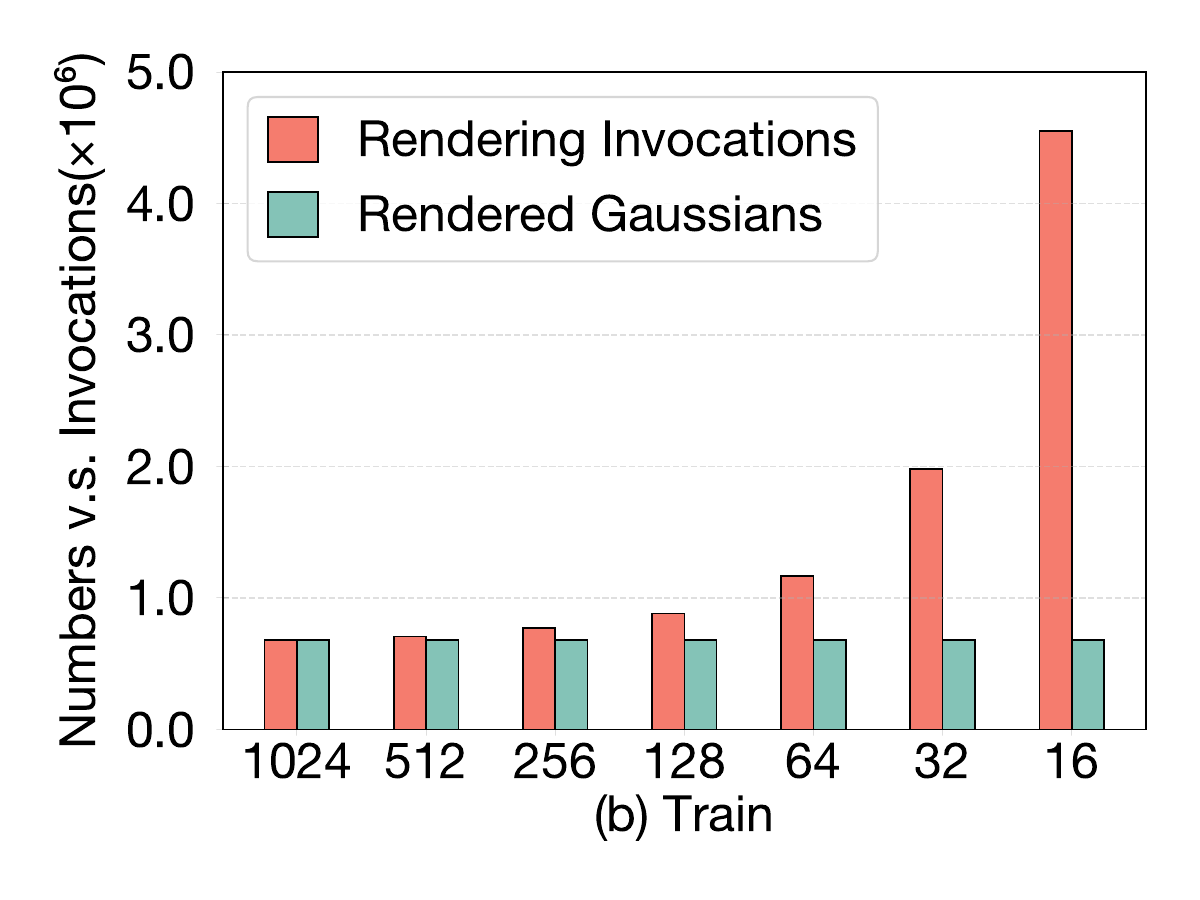}
        \vspace{-20pt}
    \caption{The overhead of Gaussian loading when partitioning an image into sub-views of size $n\times n$ in Gaussian-wise rendering.}
        \vspace{-10pt}
    \label{fig:f6}
\end{figure}

To accommodate memory-constrained edge platforms, we design a Compatibility Mode (Cmode) that partitions the full image plane into smaller sub-views using frustum tiling. Each sub-view is rendered independently, allowing the on-chip buffers to operate under tighter resource constraints without modifying the core Gaussian-wise pipeline. As shown in Figure \ref{fig:f6}, when the sub-view size is larger than 128$\times$128, Compatibility Mode introduces only marginal overhead in terms of redundant Gaussian processing compared to full-view rendering. 

\subsection{Reusing Compute Units for Grouping}
To ensure correct Gaussian-wise rendering order, the depth of all Gaussians must be computed before entering the pipeline. Since rendering in 3DGS follows a strict back-to-front strategy, unknown depths would cause incorrect scheduling and blending. Therefore, depth calculation and grouping are performed globally at the beginning of each frame as a standalone preprocessing stage, referred to as Stage I in Figure~\ref{fig:f4}.

\begin{figure}
  \centering
    \includegraphics[width=\linewidth]{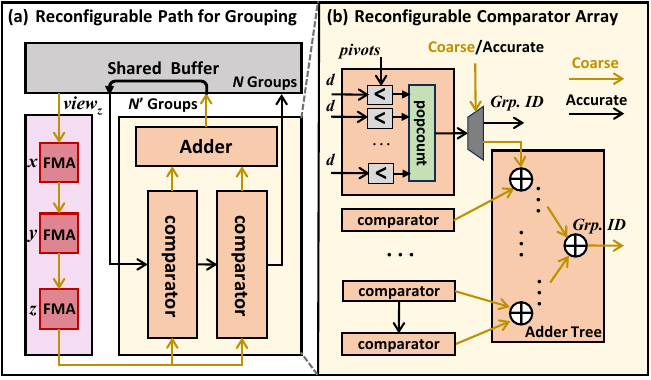}
    \vspace{-20pt}
    \caption{Grouping by shared resources.}
    \vspace{-10pt}
    \label{fig:f7}
\end{figure}

\begin{figure}
  \centering
\centering
    \includegraphics[width=\linewidth]{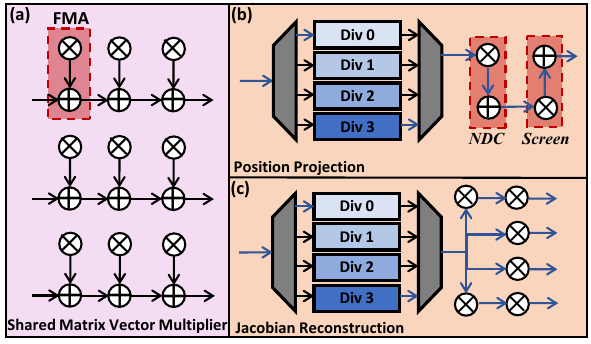}
    \vspace{-20pt}
    \caption{Key logic in Projection Unit.}
    \vspace{-10pt}
    \label{fig:f8}
\end{figure}

Importantly, both the MVM in Figure~\ref{fig:f8}(a) and the RCA in Figure~\ref{fig:f7}(b) used in Stage I are essential hardware components already required by Stage II and Stage III. Given the logical separation, our design goal is to reuse these compute and memory resources to minimize hardware cost and data movement. After grouping, the depth values and sorted IDs are stored back to DRAM via a shared on-chip buffer, preparing them for reuse in the rendering pipeline.

To compute depth efficiently, we reuse the MVM units embedded in the Projection Unit of Stage II. These units are designed to handle multiple matrix operations in Stage II. In Stage I, we instantiate 4 parallel MVMs to simulate the depth estimation. Once depth values are computed, Gaussians are passed into the RCA for hierarchical grouping. In the coarse grouping stage, the RCA operates in parallel with a cascaded adder-tree structure, enabling high-throughput comparison across millions of Gaussians. A Z-axis pivot of 0.2 is used to eliminate Gaussians outside the viewing frustum, effectively performing the first level of frustum culling. The remaining Gaussians are grouped into tens of thousands of coarse bins. For coarse bins with more than \( N \) Gaussians (i.e., \( N' > N \)), a recursive subdivision is applied to ensure that the number of Gaussians in each subgroup does not exceed the threshold \( N \), we set  \( N = 256\) in GCC.

\subsection{Projection Unit}

In Projection Unit, the PPU receives the mean vector $\mu$ of each Gaussian from the shared buffer, projects it into camera view space, and further transforms it into screen-space coordinates. Then, the RU jointly reconstructs the same Gaussian's mean $\mu$, scale $s$, and rotation $q$ to form a 3D covariance matrix and applies a perspective projection, resulting in a 2D inverse covariance matrix $\Sigma'$. Finally, the SCU estimates the elliptical coverage region of each Gaussian using $\Sigma'$ and prunes those that fall outside the viewing frustum.

In the PPU, the mean vector $\mu$ is first transformed by the view matrix into camera space. This process is performed by three parallel MVM, corresponding to the three channels of the projected position. Further transformation into NDC and pixel-space coordinates requires a division and square root operation, which introduces pipeline delays. To balance computational precision and hardware area, we implement a four-cycle iterative fused division and square root unit. Four of these units are deployed along the critical path and scheduled in an interleaved fashion, enabling one new Gaussian to be processed per cycle. This structure ensures high-throughput projection without pipeline stalls. The same unit design is reused later in the SH unit for vector norm normalization.

The RU module decodes the scale $s$ and quaternion rotation $q$ of each Gaussian to reconstruct the 3D covariance matrix $\Sigma$ and rotation matrix $R$. In parallel, the RU computes a Jacobian matrix $J$ shown in Figure~\ref{fig:f8}(c). These matrices are subsequently fed into MVMs for sequential multiplications as Equation~\ref{eq:1}.

The SCU identifies off-screen Gaussians and prunes them early in the rendering pipeline to reduce computational overhead. It receives $\Sigma'$ from the RU and estimates the elliptical coverage of each Gaussian based on its principal axes and covariance eigenvalues. To approximate the effective visible region of each Gaussian, we apply the $\omega-\sigma$ law described in Equation~\ref{eq:8}. Gaussians with no overlap are marked as invisible. The opacity $\omega$ is computed offline in log-space, and since it is view-independent for a given 3DGS model, it does not require on-chip processing and the Alpha Unit directly consumes the log-space $\omega$ values.

\subsection{Alpha Unit}

In the 3DGS rendering pipeline, each projected Gaussian must compute a transparency weight $\alpha$ for its coverage on the image plane. As shown in Equation ~\ref{eq:7}, $\alpha$ is modeled as the exponential of a negative quadratic form, derived from the pixel's relative position and the inverse covariance matrix. To efficiently support this computation in hardware, we implement a lookup-table-based EXP unit using piecewise linear approximation. Given the range of meaningful $\alpha$ values lies in $[1/255, 1)$, the corresponding exponent input is constrained to $[-5.54, 0)$. Our LUT covers only this interval; inputs below $-5.54$ are clamped to $\alpha = 0$, and those above $0$ are saturated to $\alpha = 1$. The LUT contains 16 linear segments, ensuring less than 1\% approximation error while significantly reducing area and storage overhead. Unlike GSCore~\cite{lee2024gscore}, which suffers from FP16 overflows in EXP operations due to value mapping, our design avoids such issues through fully fixed-point arithmetic in the EXP unit.

Despite early culling stages, redundant $\alpha$ evaluations persist in regions with negligible visual contribution, as illustrated in Figure~\ref{fig:f3}. To address this, we integrate a runtime Alpha-based Gaussian Boundary Identification algorithm within the Alpha Unit. This block-level identification scheme dynamically eliminates pixel clusters with insignificant alpha contributions based on the actual spatial extent of each Gaussian. The screen is divided into fixed-size $n \times n$ pixel blocks, and a corresponding $n \times n$ PE array processes each block's alpha computations in parallel. We set $n$ = 8 in GCC. For each Gaussian, block traversal starts from the pixel block containing its projected center; if the center lies outside the image bounds, traversal begins from the nearest image corner. A status map $S$ is maintained on-chip, with each block marked as unvisited initially.

Traversal proceeds in a breadth-first manner along eight octant directions. If all alpha values on the boundary of a direction fall below a threshold ($\alpha < 1/255$), the corresponding region from the current block to the image edge is marked as pruned in $S$, skipping further evaluation in that direction. This directional early termination leverages the convex support of projected Gaussians to significantly reduce unnecessary computation. During execution, pixel blocks are streamed into the compute array in the order defined by the traversal queue. Once all blocks for the current Gaussian are dispatched, the next Gaussian’s parameters are prefetched to maintain throughput. Given a per-Gaussian latency of 14 cycles, we reserve space for up to 16 Gaussians' status map S and queue Q in the on-chip preload buffer to sustain continuous pipeline operation.

\begin{figure}
  \centering
  % \captionsetup{font=small}
\centering
    \includegraphics[width=\linewidth]{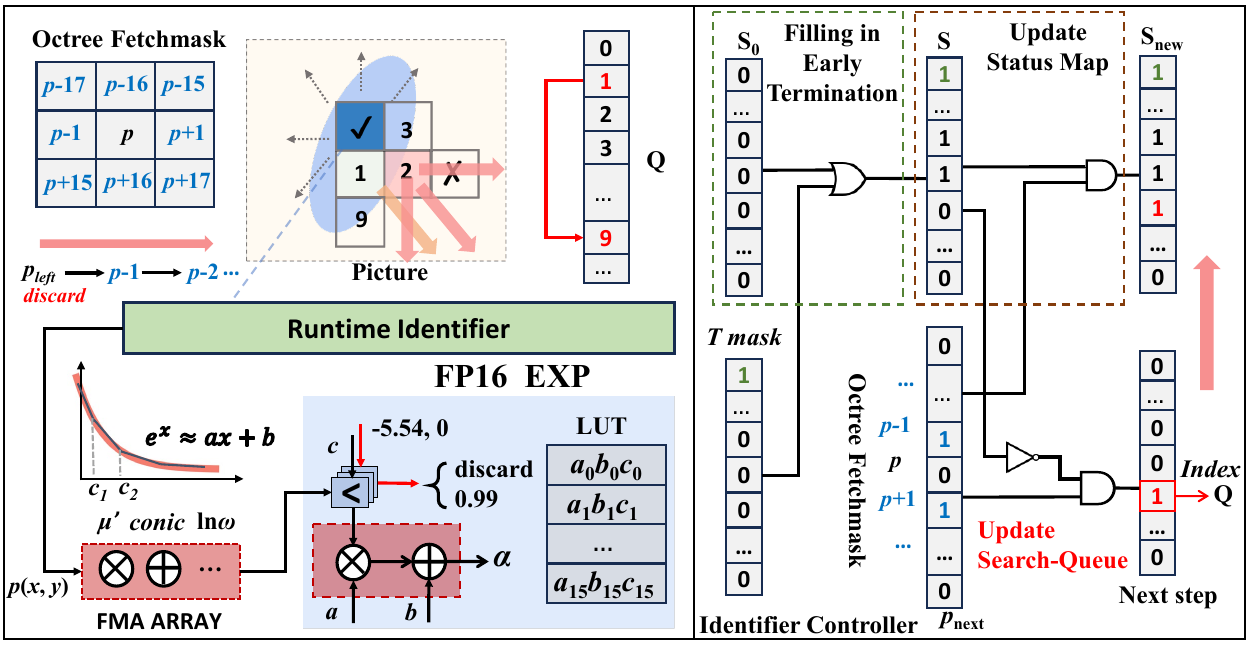}
    \vspace{-20pt}
    \caption{
    Alpha Unit and alpha-based controller.
    }
    \vspace{-10pt}
    \label{fig:f9}
\end{figure}

\subsection{Blending Unit}

After the $\alpha$ computation for a given pixel block is complete, if the block passes the transparency threshold check, its result is forwarded to the Blending Unit. The $n \times n$ array of $\alpha$ values and corresponding temporary transmittance values $T$ are processed in parallel using FMA units. The updated $T$ is then used to accumulate the intermediate RGB color values for the block, as defined in Equation \ref{eq:4}. The resulting blended outputs are written back to the Image Buffer for subsequent Gaussian compositions.

Since 3DGS requires strict back-to-front blending, the pipeline enforces ordering at the block level. If a later Gaussian in the sorted sequence attempts to access a block whose previous Gaussian has not yet completed processing, the pipeline stalls until the prior Gaussian finishes, ensuring correctness of blending results. To avoid unnecessary computation, we introduce a transmittance mask ($T_\text{mask}$) that disables blending for blocks whose $T$ value has already fallen below a predefined threshold. This mask is updated during blending and applied during the initialization of the status map $S$ for the next Gaussian. Any block with $T_\text{mask} = 1$ is excluded from future $\alpha$ computation. The updated $T$ and accumulated color values are written back to the buffer and reused in the blending of subsequent Gaussians.

\subsection{Scheduling under Compatibility Modes}

Our rendering pipeline adopts a Gaussian-wise dataflow, enabling each Gaussian to independently undergo projection, identifier, and blending to minimize redundant computation. However, on resource-constrained edge platforms, the on-chip image buffer often cannot accommodate a full-resolution frame. To address this, we design a Compatibility Mode (Cmode) that dynamically adapts the pipeline to large-scale scenes without sacrificing overall rendering efficiency.

When the hardware detects that the target image exceeds the supported resolution, the pipeline automatically switches to Cmode. In this mode, the image is partitioned into multiple sub-views of manageable size and processed sequentially. To support this, we implement a 2D spatial binning scheme, where Gaussians are grouped not only by depth but also by their screen-space coordinates. Each spatial group corresponds to a specific sub-view, ensuring that the working set remains within the limits of on-chip storage. Based on the analysis in Figure~\ref{fig:f6}, we select a sub-view size of $128\times 128$, which introduces negligible redundancy and hardware overhead while maintaining the original Gaussian distribution.

To further improve execution efficiency under Cmode, we analyze the computational load of each pipeline stage and observe that the projection unit, when operating independently, becomes memory-bound. To alleviate this, we configure the projection stage with a parallelism factor of 2. 

\section{Evaluation}\label{sec:5}

\begin{figure*}[t]
    \centering
    \includegraphics[width=0.48\textwidth]{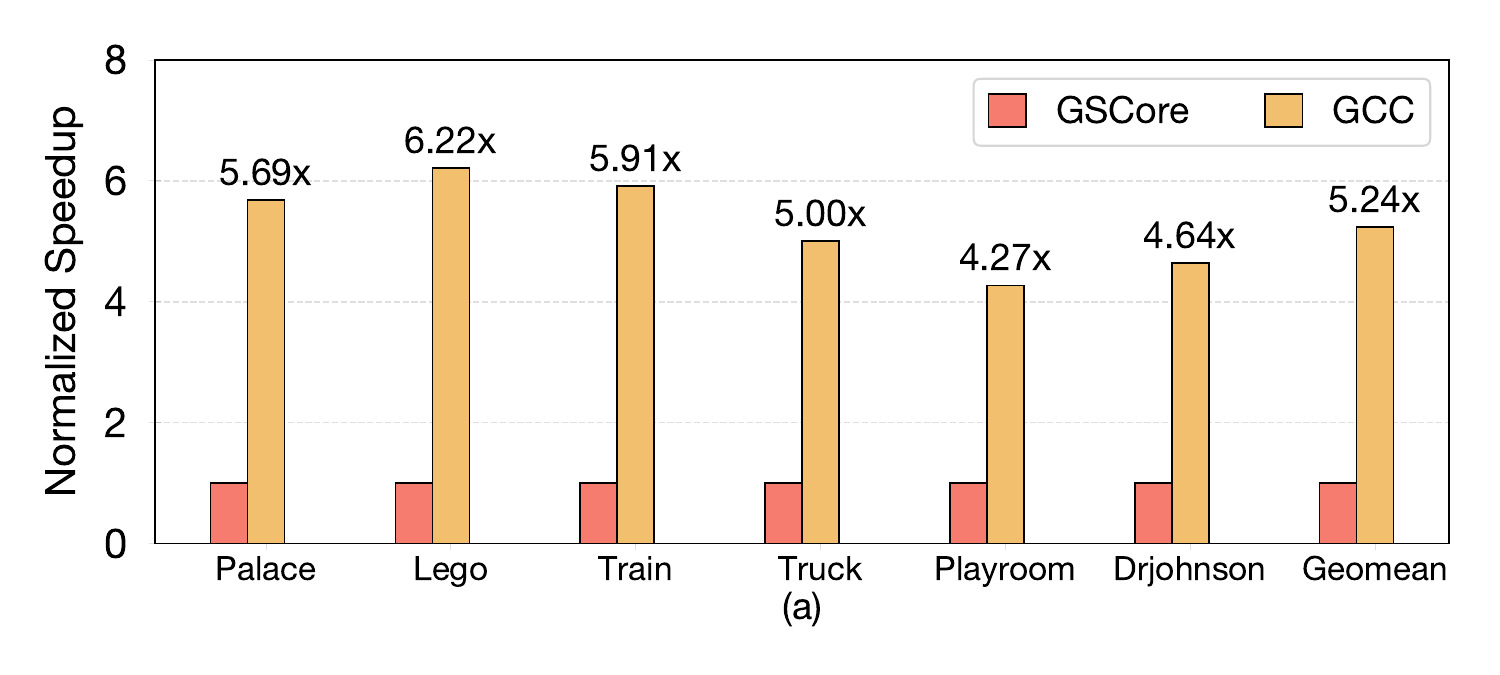}
    \hfill
    \includegraphics[width=0.48\textwidth]{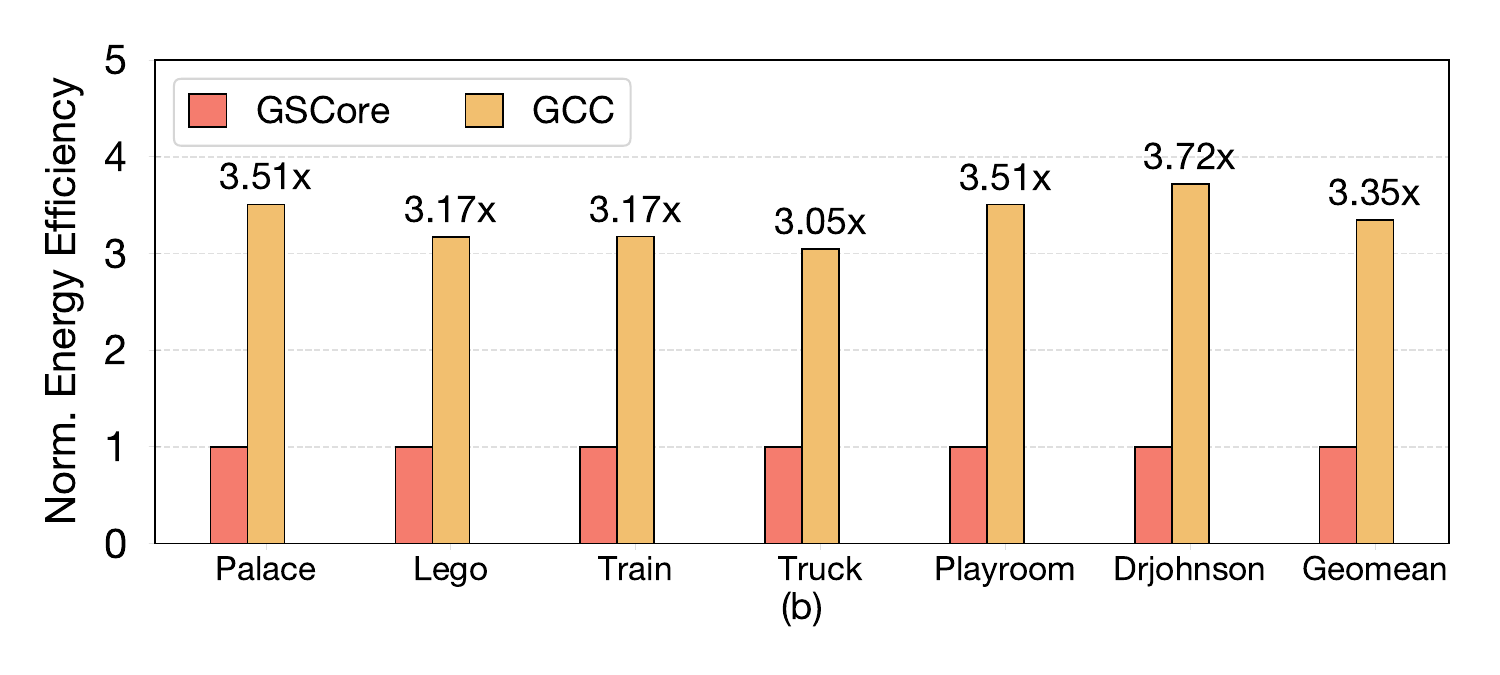}
        \vspace{-15pt}
    \caption{The area-normalized performance and energy efficiency per frame of GCC and GSCore \cite{lee2024gscore} on the six scenes.}
    \label{fig:f10}
\end{figure*}

\subsection{Methodology}

\begin{figure*}[!b]
    \centering
    \includegraphics[width=0.32\textwidth]{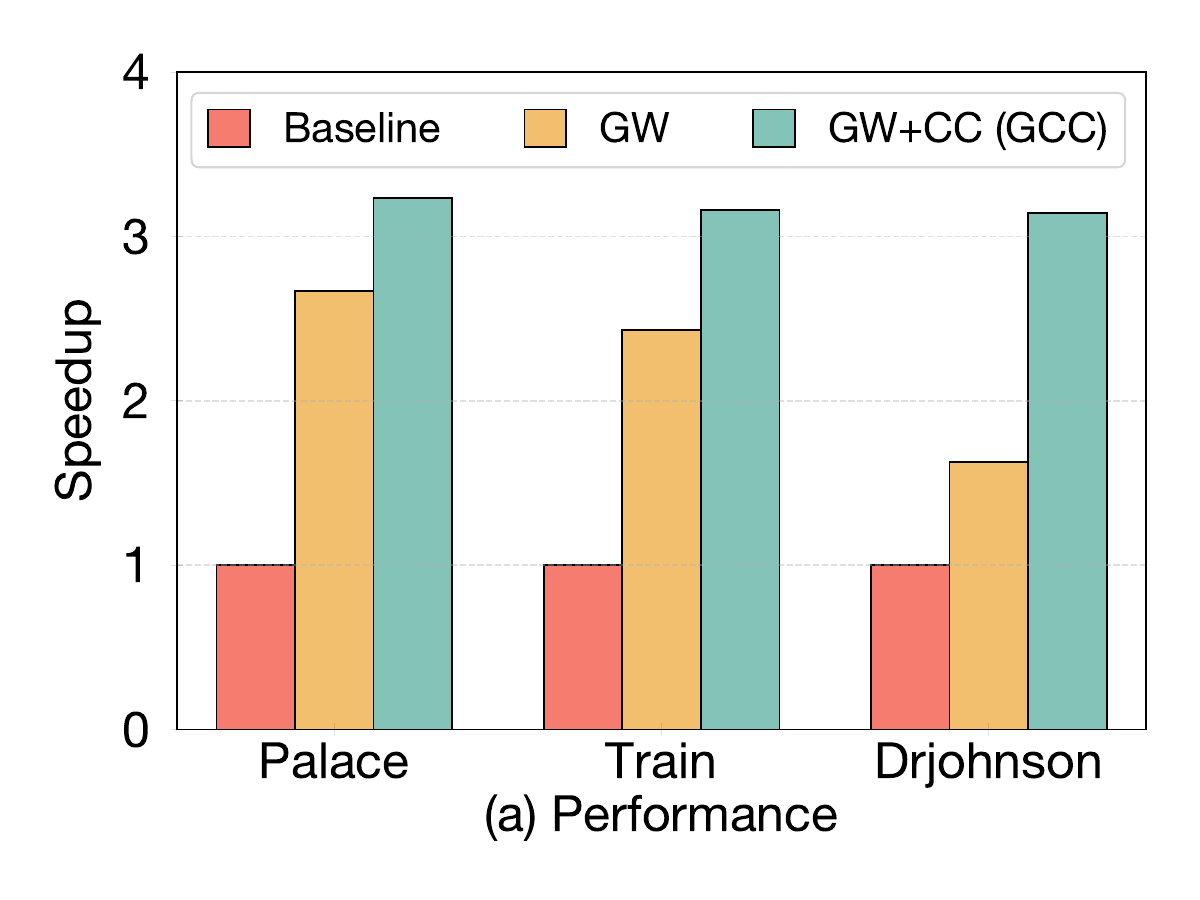}
    \hfill
    \includegraphics[width=0.32\textwidth]{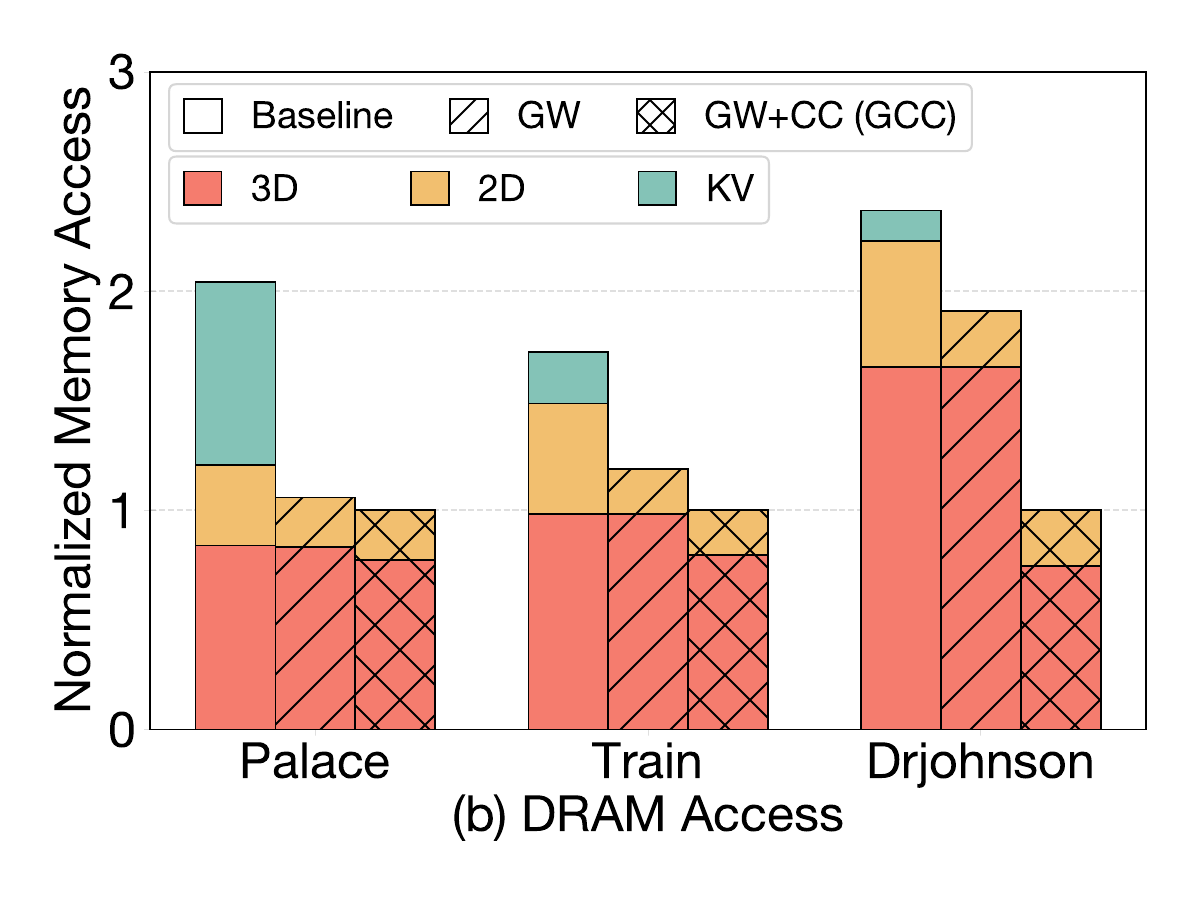}
    \hfill
    \includegraphics[width=0.32\textwidth]{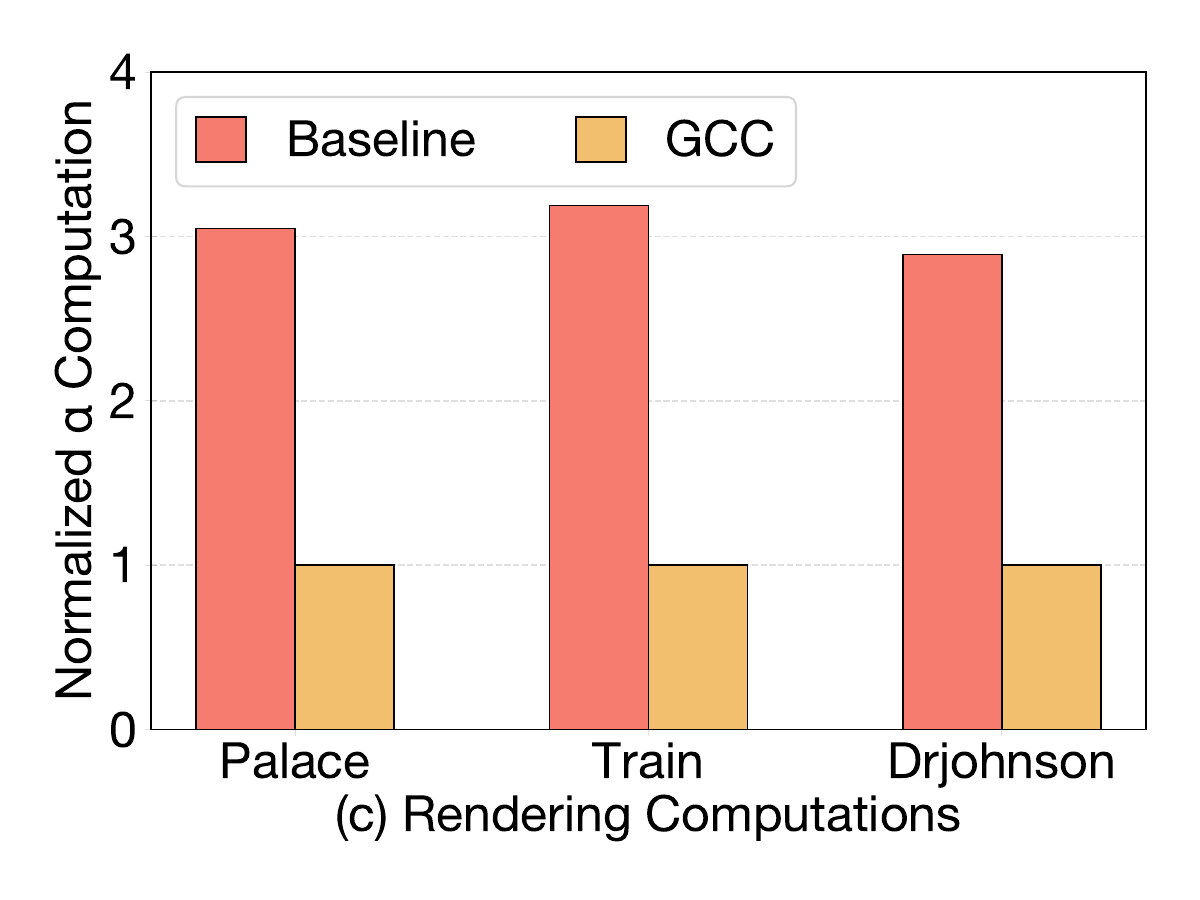}
    \vspace{-10pt}
    \caption{Breakdown Analysis of performance and DRAM access for GCC.}
    \label{fig:f11}
\end{figure*}

\textbf{Baseline and Dataset:} To demonstrate the efficiency of the proposed GCC dataflow and architecture, we select the state-of-the-art 3DGS inference accelerator, GSCore~\cite{lee2024gscore}, as our major baseline, which employs the standard two-stage processing with tile-wise rendering, and is 10$\times$ faster than an NVIDIA Xavier NX mobile GPU. 
We adopt the six representative datasets used in GSCore, covering indoor, outdoor, and synthetic scenes, to ensure consistency in scene complexity and fairness in comparison with prior work. For datasets Tanks and Temples~\cite{knapitsch2017tanks} (T\&T) and Deep Blending~\cite{hedman2018deep}  (DB), we directly use the pre-trained Gaussian models released by the original 3DGS paper. For scenes not publicly provided in the original paper, such as Lego and Palace, we follow the same training protocol and reproduce the models locally using an RTX 3090 GPU~\cite{nvidia3090} to ensure consistency in model quality. All evaluations are conducted using high-quality 3DGS models trained with 30K iterations. 

\textbf{Implementation:}
We implement the proposed GCC accelerator in SystemVerilog, and synthesize the RTL using Synopsys Design Compiler with a commercial 28\,nm standard cell library at 1\,GHz to obtain the area and power of compute units. The on-chip buffers are modeled using CACTI-P~\cite{cacti-p:iccad}. We develop a cycle-accurate simulator in Python to model both data traffic and the execution behavior of key compute units in our architecture. Each hardware module is modeled as a dedicated Python class that performs functionally-correct computations for the rendering pipeline while internally tracking the cycle-level execution cost of each operation. The execution cycles of the computation modules of all simulators have been validated with the HDL design at the cycle level.
To match the memory configuration used in GSCore, we adopt a Micron LPDDR4-3200 DRAM model as the off-chip memory system, with a peak bandwidth of 51.2\,GB/s.

For a fair comparison, we also develop a simulator for GSCore based on the architectural details provided in its paper with less than 3\% performance deviation. All evaluations are conducted under the same configuration settings as reported. Our reproduced performance results closely match the reported data, confirming the reliability of our simulator.

\begin{table}[t]
    \centering
    \caption{Comparison of rendering quality.}
    \vspace{-10pt}
    \label{tab:t4}
    \resizebox{\linewidth}{!}{%
    \begin{tabular}{cccccccc}
        \toprule
        \textbf{Method} & \textbf{Metrics}  
                               & Palace & Lego & Train & Truck & Playroom & Drjohnson \\
        \midrule
        \multirow{2}{*}{GPU}    &PSNR $\uparrow$ & 38.35 & 34.90 & 24.66 & 26.82 & 36.18 & 35.18 \\
                               &LPIPS $\downarrow$& 0.02  & 0.03 & 0.23  & 0.17  & 0.25 & 0.25 \\
        \midrule
        \multirow{2}{*}{GSCore} &PSNR $\uparrow$& 38.36 & 34.90 & 24.70  & 26.9  & 36.62 & 35.32 \\
                               &LPIPS $\downarrow$& 0.02  & 0.03 & 0.23  & 0.17  & 0.25 & 0.25 \\
        \midrule
        \multirow{2}{*}{GCC}   &PSNR $\uparrow$ & 38.35 & 34.90 & 24.70  & 26.9  & 36.49 & 35.29 \\
                               &LPIPS $\downarrow$& 0.02  & 0.03 & 0.23  & 0.17  & 0.25 & 0.25 \\
        \bottomrule
    \end{tabular}
    \vspace{-20pt}
    }
    \begin{flushleft}
    \footnotesize
    \end{flushleft}
    
\end{table} 

\subsection{Performance and Quality}
\begin{table}[t]
  \centering
  \caption{Comparison of Neural Rendering Accelerators}
  \vspace{-10pt}
  \label{tab:t6}
  \resizebox{\linewidth}{!}{%
  \begin{tabular}{>{\centering\arraybackslash}p{2cm}|>{\centering\arraybackslash}p{1.6cm}>{\centering\arraybackslash}p{1.6cm}>{\centering\arraybackslash}p{1.6cm}>{\centering\arraybackslash}p{1.6cm}>{\centering\arraybackslash}p{1.6cm}>{\centering\arraybackslash}p{1.6cm}}
    \toprule
    & \textbf{MetaVRain ISSCC'23} & \textbf{Fusion-3D MICRO'24} & \textbf{Nvidia A6000} & \textbf{Jetson AGX Xavier} & \textbf{GSCore ASPLOS'24} &\textbf{\makebox[0.2cm]{}GCC\hspace{0pt}\makebox[0.2cm]{}} \textbf{This Work}\\
    \midrule
    \textbf{Model} & NeRF & NeRF & 3DGS & 3DGS & 3DGS & 3DGS \\
    \midrule
    \textbf{PSNR} & 31.58dB & 31dB & 35.78dB & 35.54dB & 34.90dB & 34.90dB \\
    \midrule
    \textbf{Process} & 28nm& 28nm & 8nm & 12nm & 28nm & 28nm \\
    \midrule
    \textbf{Area} & 20.25mm\textsuperscript{2} & 8.7mm\textsuperscript{2} & 628mm\textsuperscript{2} & 350mm\textsuperscript{2} & 3.95mm\textsuperscript{2} & \textbf{2.71mm\textsuperscript{2}} \\
    \midrule
    \textbf{SRAM} & 2015KB & 1099KB & -- & -- & 272KB & \textbf{190KB} \\
    \midrule
    \textbf{Frequency} & 250MHz & 600MHz & 1040MHz & 854MHz & 1GHz & 1GHz \\
    \midrule
    \textbf{Power} & 0.89W & 6.0W & 300W & 30W  & 0.87W & \textbf{0.79W} \\
    \midrule
    \textbf{Throughput$^{*}$} & 110FPS& 36FPS & 300FPS &  20FPS & 190FPS & \textbf{\textbf{667FPS}} \\
    \midrule
    \textbf{Area-normed Throughput} & 5.43 FPS/mm\textsuperscript{2}& 4.13 FPS/mm\textsuperscript{2} & 0.48 FPS/mm\textsuperscript{2} & 0.05 FPS/mm\textsuperscript{2}  & 48.10 FPS/mm\textsuperscript{2} & \textbf{246.00 FPS/mm\textsuperscript{2}} \\
    \bottomrule
  \end{tabular}
  \vspace{-10pt}
  }
  \raggedright\small$^{*}$All throughput values are tested on the Lego dataset, as prior NeRF works such as Fusion-3D focus on the Synthetic-Nerf benchmark.
  
\end{table}
To ensure that these performance gains do not compromise rendering fidelity, Table~\ref{tab:t4} compares image quality metrics—including PSNR and LPIPS~\cite{zhang2018unreasonable}—across GCC, GSCore, and the original GPU implementation. All methods produce nearly indistinguishable visual results, with PSNR deviations below 0.1 dB and identical LPIPS scores, indicating imperceptible quality degradation.

Figure~\ref{fig:f10}(a) illustrates the area-normalized throughput across six benchmark scenes. GCC consistently outperforms GSCore on all datasets, achieving speedups ranging from 4.27$\times$ (Playroom) to 6.22$\times$ (Lego), with a geometric mean of 5.24$\times$. These improvements stem from GCC’s Gaussian-wise rendering and cross-stage conditional execution, which effectively eliminate redundant computations and reduce memory traffic compared to the conventional tile-based pipeline.

To further attribute performance gains to specific architectural innovations, Figure~\ref{fig:f11}(a) illustrates an ablation study across three representative scenes: Palace (compact synthetic scene), Train (medium-scale real scene), and Drjohnson (large-scale real scene).
Combined with Figure~\ref{fig:f10}(a), the results confirm that our two key contributions, Gaussian-wise (GW) rendering and cross-stage conditional (CC) execution, consistently deliver performance benefits across scenes of varying complexity and scale.

For compact scenes such as Palace, where most Gaussians cluster near the camera center, GW is especially effective in reducing tile-level redundancy. In these cases, the Alpha-based Identifier further improves runtime by aggressively pruning unnecessary blending operations. In contrast, large-scale scenes like Drjohnson exhibit sparse Gaussian distributions. Here, CC plays a more prominent role by suppressing the SH evaluation and projection of visually ineffective Gaussians. Figure~\ref{fig:f11}(b) quantifies reductions in off-chip memory accesses for 3D Gaussian attributes, 2D projected ellipses, and tile-level key-value (KV) mappings. GW rendering eliminates repeated KV lookups, while CC processing avoids loading Gaussians that do not contribute to final rendering results. Figure~\ref{fig:f11}(c) confirms that Alpha-based pruning contributes to computation reduction across all scene types.

Table~\ref{tab:t6} further compares GCC with recent neural rendering accelerators, including MetaVRain \cite{han2023metavrain} and Fusion-3D \cite{li2024fusion} for NeRF, and GSCore for 3DGS. While MetaVRain and Fusion-3D target implicit neural fields and differ in rendering characteristics, they represent prominent dedicated accelerators in neural scene rendering.
We also include the NVIDIA A6000~\cite{nvidiaA6000} and NVIDIA Jetson AGX Xavier \cite{nvidia_xavier_2018} GPUs as the high-end and low-end references, respectively.
Compared to NeRF accelerators, GCC achieves more than 40$\times$ higher area-normalized throughput while requiring 5$\times$/10$\times$ less SRAM, under similar PSNR levels.
Compared to GSCore, GCC achieves 3.5$\times$ throughput, 30\% lower area.
Through its Gaussian-wise execution model and cross-stage optimizations, GCC delivers GPU-level throughput with <1W power and <0.3\% of the A6000’s silicon area, demonstrating strong efficiency and scalability under edge constraints.

\subsection{Area and Energy Efficiency}
 
Table~\ref{tab:t5} presents the area and power breakdown of our GCC architecture. To account for architectural differences from GSCore, we separately report contributions from compute units and on-chip buffers, with proportional breakdowns shown in parentheses. While both designs employ identical numbers of FPEXP modules, GCC reduces the number of FPMACs without sacrificing throughput. This is enabled by our balanced Gaussian-wise dataflow, which leverages CC scheduling to lower the required parallelism in both compute and memory subsystems—reducing SH evaluation to 1-way and projection to 2-way, as opposed to the 4-way parallelism in GSCore. Overall, GCC occupies 40\% less area compared to GSCore, while exhibiting marginally lower power consumption.

\begin{figure}
  \centering
\centering
    \includegraphics[width=\linewidth]{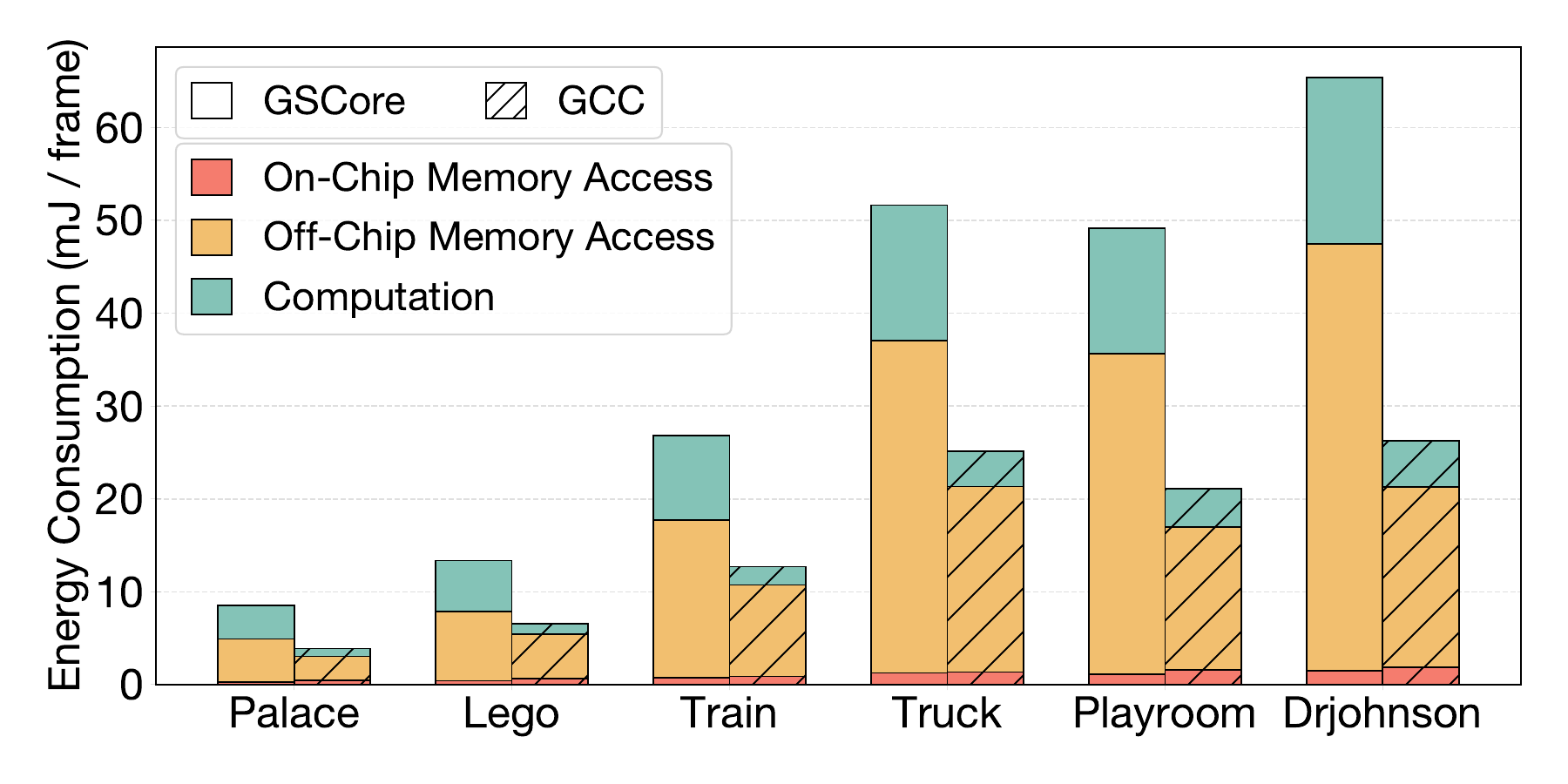}
    \vspace{-20pt}
    \caption{
 Energy breakdown of GCC and GSCore \cite{lee2024gscore}.
    }
    \vspace{-10pt}
    \label{fig:f12}
\end{figure}

Figure~\ref{fig:f12} provides the per-frame energy consumption breakdown. In both designs, DRAM accesses dominate overall energy usage. However, by avoiding repeated loading of the same Gaussian and eliminating tile-induced redundancy, GCC reduces DRAM traffic by over 50\%. While SRAM activity slightly increases due to intensified data exchange between the Blending Unit and the Image Buffer under the Gaussian-wise pipeline, the substantially lower energy cost per SRAM access compared to DRAM results in significant overall energy savings.
By significantly reducing the alpha computation and blending workload, alpha-based identifier accelerates per-frame execution and consequently lowers the dynamic energy consumed during each frame.

From a system-level perspective, GCC converts high-cost, repeated DRAM accesses into low-cost, one-pass off-chip interactions. This reorganization of memory access patterns, combined with reduced per-frame latency and minimized data movement, enables GCC to operate with high energy efficiency under stringent power constraints. As shown in Figure~\ref{fig:f10}(b), GCC achieves a 3.35$\times$ improvement in area-normalized energy efficiency over GSCore across all benchmark scenes.

\begin{table}[t]
\caption{Area and Power Comparison with GSCore.}
\vspace{-10pt}
\label{tab:t5}
\centering
\resizebox{\linewidth}{!}{
\begin{tabular}{cccc}
\toprule
\textbf{Component} & \textbf{Area (mm²)} & \textbf{Power (mW)} & \textbf{Configuration} \\ \midrule
    RCA & 0.010 & 2 & 4 units\\ 
    Projection Unit & 0.358 & 147 & 2 units\\ 
    SH Unit & 0.339 & 141 & 1 units \\ 
    Sorting Unit & 0.010 & 11 & 1 units\\ 
    Alpha Unit & 0.576 & 266 & 64 PEs \\ 
    Blending Unit & 0.382 & 172 & 64 PEs \\  \midrule
    Total & 1.675 & 739 & -\\ \midrule
    Shared Buffer & 0.019 & 3 & $2\times1\times6$~KB \\ 
    SH Buffer & 0.116 & 10 & $2\times3\times8$~KB \\ 
    Sorted Buffer & 0.029 & 1 & $2\times1\times1$~KB \\ 
    Image Buffer & 0.872 & 37 & $1\times4\times32$~KB \\ \midrule
    Total & 1.036 & 51 & 190~KB\\ \midrule
    \textbf{GCC Total} & \textbf{2.711} & \textbf{790} & -\\ \midrule \midrule
    Compute Units & 2.70 & 830 & -\\
    Buffer & 1.25 & 40 & 272~KB\\ \midrule
    \textbf{GSCore Total} & \textbf{3.95} & \textbf{870} & -\\ \midrule
\bottomrule
\end{tabular}
\vspace{-10pt}
}
\end{table}

\subsection{Sensitivity Study}

\begin{figure}[t]
    \centering
    \includegraphics[width=0.235\textwidth]{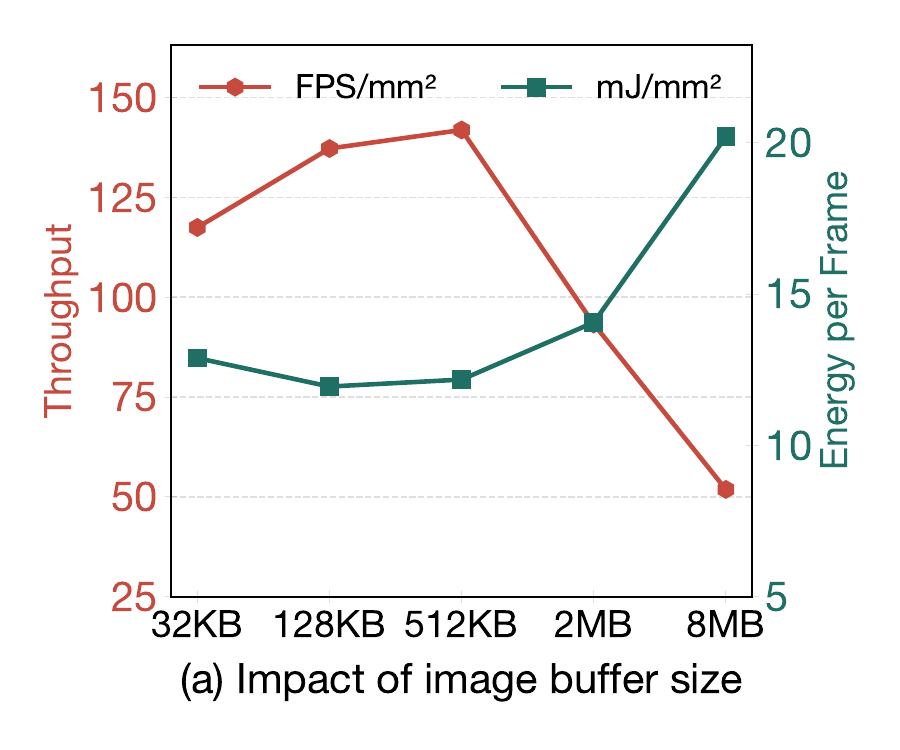}
    \hfill
    \includegraphics[width=0.235\textwidth]{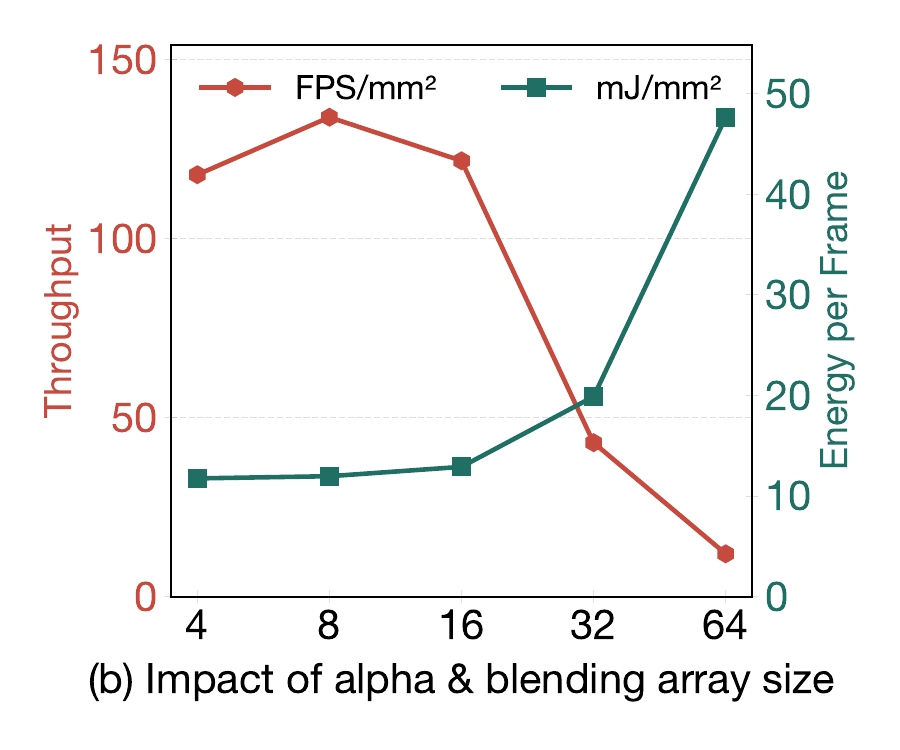}
    \vspace{-20pt}
    \caption{Design space exploration on performance-per-area (FPS/mm²) and energy-per-area (mJ/mm²) under different architectural configurations evaluated for the Train scene.}
    \vspace{-10pt}
    \label{fig:f13}
\end{figure}

\textbf{Image Buffer Size.} 
The image buffer is a critical design factor that determines the scalability and edge deployability of the GCC architecture. On resource-constrained platforms such as edge devices, the on-chip buffer capacity directly determines whether full Gaussian-wise + Cross-stage Conditional (CC) rendering can be supported. Based on the analysis in Figure~\ref{fig:f6}, we adopt a $128\times128$ image buffer combined with compatibility mode to support complex scenes efficiently. Notably, rendering accuracy remains unchanged across different sub-view sizes. This is an expected outcome, as our compatibility mode only alters the processing order of Gaussians. As demonstrated in earlier sections, this configuration achieves strong performance while maintaining reasonable area.
To further validate this design choice, we explore the design space from a performance perspective. As shown in Figure~\ref{fig:f13}(a), configurations with 128KB and 512KB buffers yield comparable normalized performance. While a larger buffer reduces runtime latency, the area overhead amortization becomes less favorable, leading to a net decline in area-normalized performance. We select 128KB as the default image buffer size to strike a balance between performance and energy efficiency.

\textbf{Alpha \& Blending Array Size.} 
This component reflects a trade-off between compute parallelism and redundancy control. Larger arrays enable higher throughput by processing more Gaussians in parallel, but they may introduce additional redundancy due to coarser boundary checks. In contrast, smaller arrays provide finer-grained control of Gaussian coverage but sacrifice overall throughput.
We explore this trade-off through design space analysis. As shown in Figure~\ref{fig:f13}(b), an $8\times8$ array provides the best performance when normalized by area. Although slightly larger arrays (e.g., $16\times16$) can reduce rendering time, the resulting increase in DRAM access latency becomes the dominant performance bottleneck and offsets the gains from additional parallelism.

\begin{figure}
  \centering
\centering
    \includegraphics[width=\linewidth]{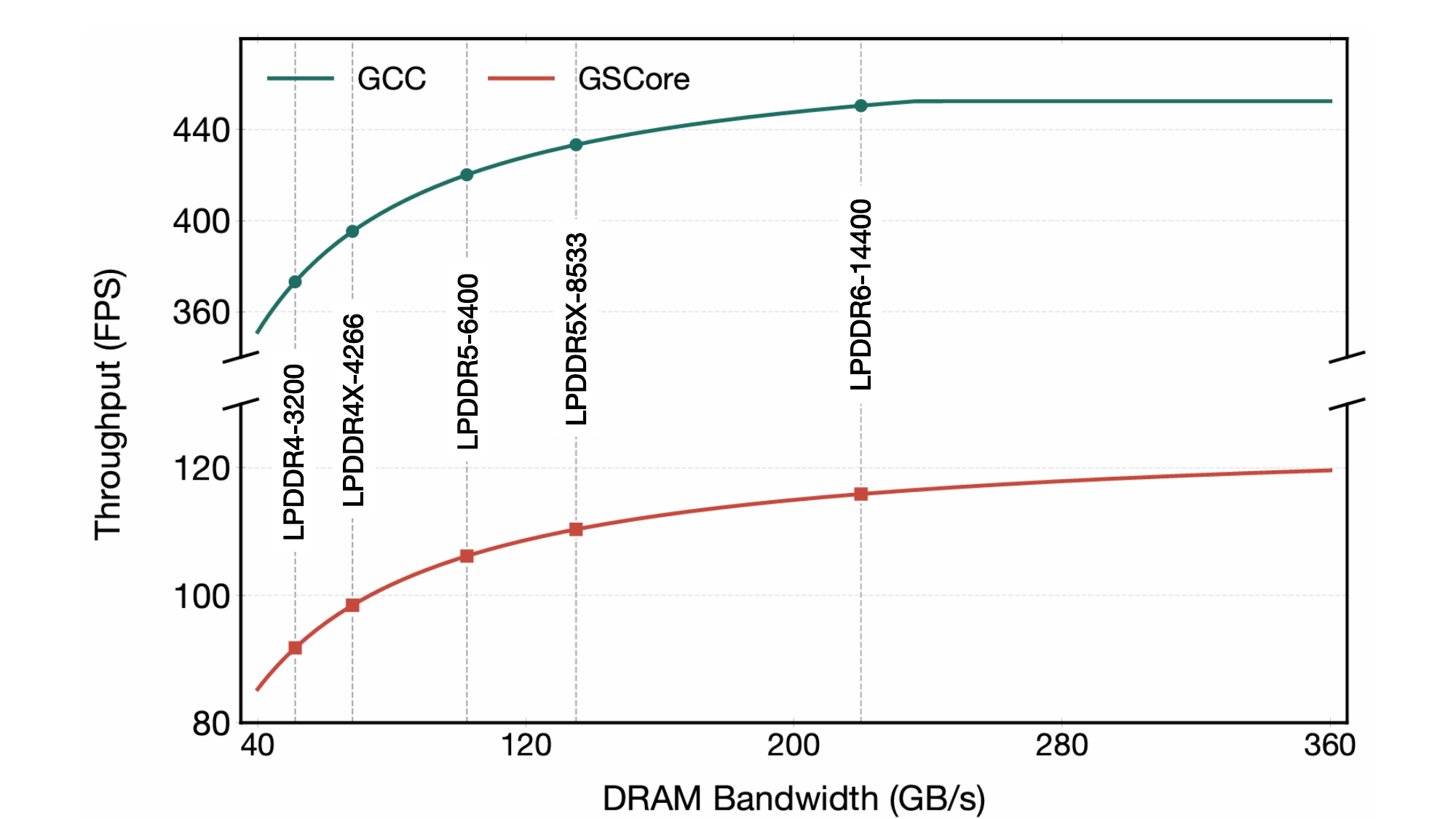}
    \vspace{-20pt}
    \caption{
Throughput comparison under different DRAM bandwidth levels for the Train scene.
    }
    \vspace{-10pt}
    \label{fig:r2}
\end{figure}

\textbf{DRAM Bandwidth.}
In the performance evaluation in Section 5.2, our GCC accelerator initially adopts the same off-chip memory configuration as GSCore \cite{lee2024gscore}—LPDDR4 with a peak bandwidth of 51.2 GB/s. In this experiment, we further investigate how varying off-chip bandwidth affects GCC's performance and compare with GSCore.
The results in Figure~\ref{fig:r2} clearly show that: 1) When DRAM bandwidth remains below 220 GB/s, both GCC and GSCore exhibit significant performance improvements as bandwidth increases. The primary reason is that the preprocessing and rendering stages in 3DGS inference require extensive on-chip/off-chip data transfers. Higher DRAM bandwidth effectively mitigates memory-bound bottlenecks. This suggests that upgrading from LPDDR4 to more advanced memory technologies (e.g., LPDDR4X, LPDDR5, LPDDR5X, or future LPDDR6 variants) would further enhance inference performance; 2) as the DRAM bandwidth exceeds 220GB/s, the performance of GSCore continues to improve slightly, but the performance of GCC remains unchanged. This is because, due to cross-stage conditional processing and Gaussian-wise rendering, GCC's off-chip data transfer volume is significantly lower than that of GSCore. As bandwidth increases beyond 220GB/s, GCC becomes compute-bound, unlike GSCore, which remains memory-bound.

\section{Discussion}

To enable efficient 3DGS inference on edge devices for wearable applications, we propose the GCC dataflow and its specialized architecture in this work. However, a noteworthy question arises: \textit{Can we satisfy the requirements for 3DGS inference in edge computing scenarios by implementing the GCC dataflow directly on GPUs, thereby obviating the need for custom hardware?}

\begin{figure}
  \centering
    \includegraphics[width=\linewidth]{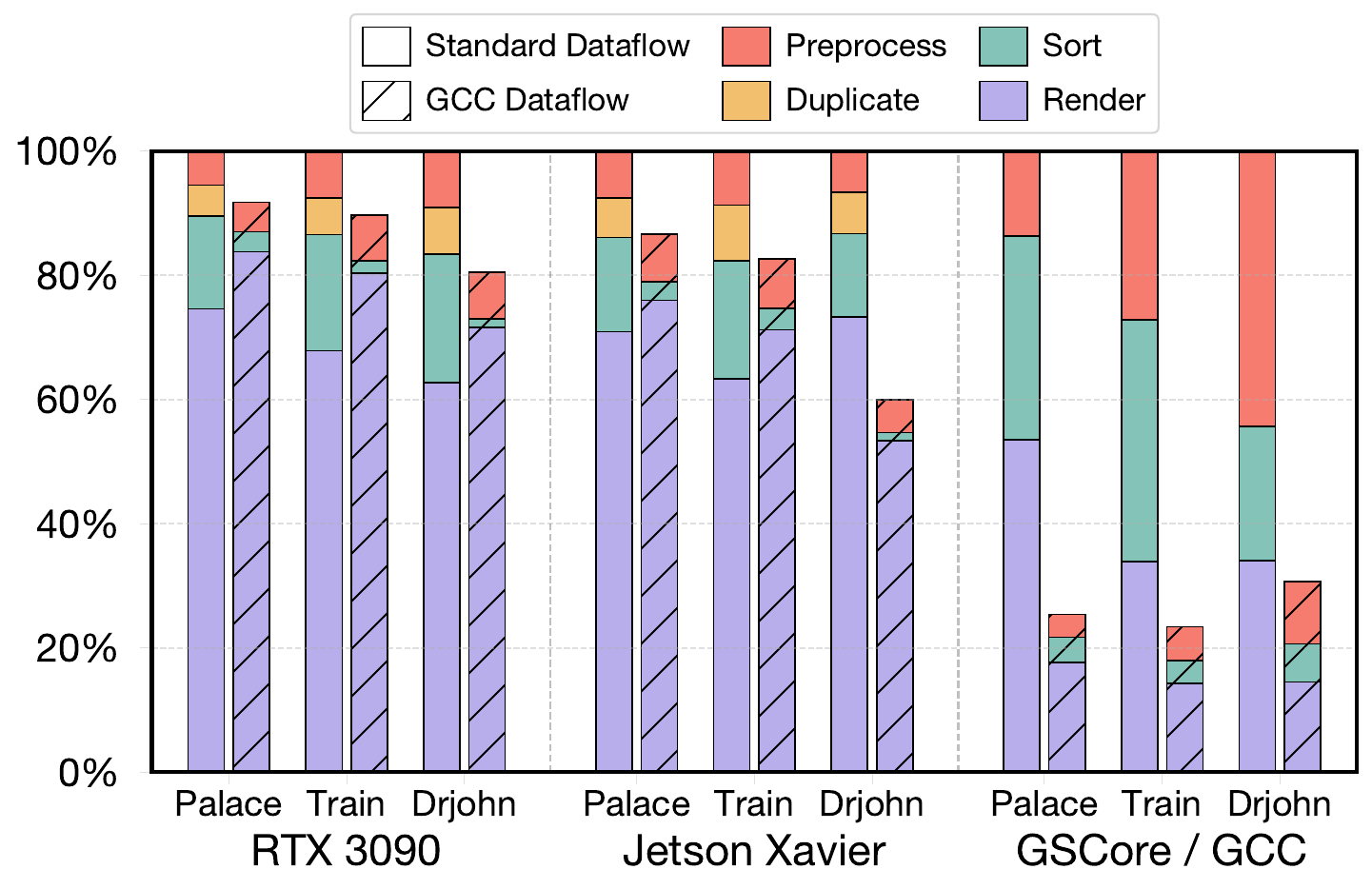}
    \vspace{-20pt}
    \caption{Breakdown analysis of per-frame execution time for the standard dataflow and our GCC dataflow on GPUs and accelerators. All results are normalized to that of the standard dataflow in each setting.
    }
    \vspace{-10pt}
    \label{fig:r1}
\end{figure}

To comprehensively evaluate the effectiveness of GCC dataflow on GPUs, we conducted experiments using both cloud-oriented (NVIDIA RTX 3090) and mobile-oriented (NVIDIA Jetson AGX Xavier) GPU platforms. For both the standard dataflow and GCC dataflow implementations, we use the PyTorch framework and apply equal optimization efforts to ensure fair comparison. Results are shown in Figure~\ref{fig:r1}.

Beyond this result, we identify two noteworthy observations. \textbf{First}, rendering time dominates the overall execution time in GPU-based 3DGS inference, making methods like GCC—which primarily target reducing redundant data movement—offer relatively limited performance improvements. This is primarily due to the abundant on-chip cache of GPUs, making 3DGS inference compute-bound. Conversely, in customized architectures with limited on-chip storage (such as GSCore \cite{lee2024gscore} with 272 KB on-chip SRAM), data movement has a more significant impact on inference performance, and as the complexity of the scenario increases (e.g., Drjohnson), the proportion of time spent on data movement becomes even greater. In such cases, our GCC dataflow demonstrates a clear advantage over standard dataflow in improving performance. \textbf{Second}, although GCC reduces rendering computation by identifying more compact Gaussian regions, the rendering time actually increases. The reason is that GCC dataflow is ``Gaussian-parallel” in our GPU implementation, where a single thread writes a Gaussian's contribution to multiple pixels. The ``many-to-one” pattern causes data races, forcing costly atomic operations for deterministic blending, which disrupts the GPU's parallel efficiency. However, our GCC accelerator can fully translate the reduction in rendering computation into savings in inference time through carefully designed compute units and pipeline parallel processing methods.

Note that GCC-based 3DGS inference on the Jetson AGX Xavier delivers just 6$\sim$20 FPS in this experiment, well below the widely recognized 90 FPS target \cite{kerbl20233d,huang2025seele}. 
By designing dedicated architectures instead of directly implementing on existing GPU platforms, we can fully unleash the performance potential of GCC dataflow, achieving inference performance ($>$150 FPS) comparable to—or even surpassing—high-end GPUs. \\

\section{Related Work}\label{sec:6}

\textbf{Neural Rendering Accelerators.} The rapid advancement of neural rendering techniques has driven the demand for real-time, high-fidelity rendering on dedicated hardware platforms. NeRF-based methods typically rely on multilayer perceptron networks to perform point-wise inference over densely sampled spatial coordinates. This leads to intensive computation and complex data dependencies, making real-time execution challenging. Consequently, a wide range of studies have proposed systematic optimizations for NeRF rendering, including algorithmic restructuring, feature encoding, and computation graph-level enhancements~\cite{muller2022instant, li2022rt, zhao2023instant, lee2023neurex, mubarik2023hardware, li2023instant, ryu202420, park202420, chen2023mobilenerf}. Recent state-of-the-art NeRF acceleration frameworks focus on improving end-to-end inference efficiency, reducing power consumption, and enabling deployment on edge devices. MetaVRain~\cite{han2023metavrain} introduces a low-power neural 3D rendering processor tailored for mobile platforms, integrating a visual perception core with hybrid neural engines and an efficient positional encoding module to support high-throughput, end-to-end NeRF inference. Fusion-3D~\cite{li2024fusion} proposes a full-stack acceleration solution combining dynamic scheduling, mixed-precision execution, and chiplet-based Mixture-of-Experts (MoE) architecture, significantly improving both training and inference performance and scalability.

\textbf{3DGS System-Level Optimization and Specialized Accelerators.} Recent works have accelerated 3DGS through both specialized hardware and system-level optimizations. Dedicated hardware efforts include ASICs targeting inference~\cite{lee2024gscore}, dynamic SLAM~\cite{wu2024gauspu}, and training~\cite{10946749}, as well as FPGA-based optimizations for pipeline stages like projection~\cite{shimamura2025exploring}. Concurrently, VR-Pipe~\cite{lee2025vr} enhance performance on existing GPUs by integrating early alpha termination into the rendering pipeline.

While some algorithmic optimizations also leverage alpha values to reduce computation~\cite{wang2024adr,hanson2025speedy,feng2025flashgs}, they operate by pruning Gaussian-tile pairings within the conventional dataflow. Beyond splatting-style rasterization~\cite{glassner1989introduction}, other works have even explored alternative rendering techniques like ray tracing for Gaussian scenes~\cite{moenne20243d}, highlighting the representation's versatility.

Despite this diversity of approaches, a common thread persists: most prior work adheres to the traditional, decoupled tile-wise rendering paradigm. This model, while modular, introduces structural inefficiencies—such as redundant memory transfers and poor data reuse—which are particularly detrimental on resource-constrained platforms. In contrast, our work fundamentally departs from this paradigm by introducing a holistic, Gaussian-wise execution model with cross-stage conditional awareness. This architecture directly targets the aforementioned inefficiencies, enabling significant gains in throughput and energy efficiency for 3DGS inference.

\section{Conclusion}\label{sec:7}
This paper introduces GCC, a specialized accelerator for 3D Gaussian Splatting (3DGS) inference. By addressing the inherent inefficiencies of tile-wise rendering and decoupled-staged execution in standard dataflow, GCC employs Gaussian-wise rendering and a cross-stage conditional processing scheme to eliminate redundant data movements and computations. The resulting architecture achieves substantial improvements in throughput and energy efficiency over prior art, while maintaining competitive rendering quality and inference latency comparable to high-end GPUs—all within a significantly smaller area and power budget.

\begin{acks}
This work was supported in part by National Key R\&D Program of China (No.2022ZD0160304), Beijing Natural Science Foundation (No.4254088), and AMD Heterogeneous Accelerated Compute Clusters (HACC) program.
\end{acks}

%%%%%%% -- PAPER CONTENT ENDS -- %%%%%%%%

\bibliographystyle{ACM-Reference-Format}
\bibliography{sample-base}

\end{document}